\pdfoutput=1  

\documentclass[sigconf,nonacm]{aamas}
\setcopyright{none}
\acmConference[Preprint]{arXiv preprint}{October 2026}{arXiv}{}
\copyrightyear{2026}
\acmYear{2026}
\acmDOI{}
\acmISBN{}

\usepackage{array}     
\usepackage{tikz}      
\usetikzlibrary{arrows.meta}
\usepackage{pdfpages}  

\newif\ifarxiv
\arxivtrue

\author{Burak Agachan}
\affiliation{%
  \institution{Leiden Institute of Advanced Computer Science (LIACS), Leiden University}
  \city{Leiden}
  \country{The Netherlands}}

\author{Max van Duijn}
\affiliation{%
  \institution{Leiden Institute of Advanced Computer Science (LIACS), Leiden University}
  \city{Leiden}
  \country{The Netherlands}}

\author{Amirhossein Zohrehvand}
\affiliation{%
  \institution{Leiden Institute of Advanced Computer Science (LIACS), Leiden University}
  \city{Leiden}
  \country{The Netherlands}}

\title[Loop-Back Authority in LLM Agent Teams]{Loop-Back Authority in LLM Agent Teams: A Paired Experiment on Flat and Hierarchical Coordination}

\begin{abstract}
Does authority in AI teams improve the outcome? Organizational theory asserts that authority facilitates decision making, improving quality. Meanwhile, some nascent AI research suggests that revision under authority makes LLM output worse. Multi-agent LLM frameworks default to giving a Manager agent the authority to send a worker's output back for revision. Prior comparisons test the effect of authority using verifiable tasks. We conduct an experiment on an open-ended task, business-intelligence reporting, using a sample of 43 paired laptop products and 86 runs. Each report is written once by a hierarchical team and once by a flat team. We find that flat teams produce higher-quality reports, scoring higher on Utility ($d = 0.42$, $p = 0.009$) and Writing Clarity ($d = 0.34$, $p = 0.030$). The reports are the same length, but hierarchical team reports use 53\% more hedging words such as ``may'' and ``could'', and each revision is associated with a 0.14-point drop in Writing Clarity on a 1 to 5 scale. Before any revision, the hierarchical team's first draft is indistinguishable from the flat team's report. In other words, the quality gap can be traced to revision. Authority improves quality when the Manager can verify the work, else when it can only provide feedback it has a negative effect on quality.
\end{abstract}

\keywords{Multi-agent systems; Organizations and institutions; Coordination; Authority and hierarchy; LLM-based agents; LLM-as-a-judge evaluation}

\begin{document}

\maketitle


\section{Introduction}

Multi-agent LLM systems are increasingly deployed in hierarchical frameworks, where a Manager agent sits above worker agents, reviews their output, and can send it back \cite{hong2023metagpt, wu2023autogen, zhang2025agentorchestra, tran2025multiagent}. The intuition is that authority improves quality as it is the case in classic organizational coordination theory, i.e., authority leads to faster convergence in decisions \cite{kocak2022dual}. A recent body of evidence suggests the opposite. RLHF-trained language models comply sycophantically with authoritative prompts \cite{sharma2024sycophancy, wei2023simple, bai2022constitutional} and lose divergent reasoning under iterative critique, a pattern called Degeneration-of-Thought \cite{liang2024encouraging}. These findings suggest that authority in the case of LLMs might produce worse output, as opposed to the classical organizational design theory. This study provides a controlled test of the two predictions of these two theories.

Multi-agent systems research has primarily conceived organizational design as an engineering variable \cite{smith1980contractnet}, settling on the position that no organizational form is best across environments \cite{horling2004survey}. What has changed is the agent. In classical organizational MAS, how a component responded to authority was fixed in its code. An authority link sent a task down and got a result back. In an LLM team, a Manager's instruction does two things. It passes a task to the worker, and it changes the text the worker writes. Therefore, we cannot assume that results from classical multi-agent systems still hold \cite{xian2025reliable}. A related line of work asks the mirror question: can an LLM manage a team of humans? In a pre-registered experiment, Zohrehvand et al.\ \cite{zohrehvand2026coordination} find that an AI manager coordinating human workers through communication alone raises team performance. The present study asks the complementary question, namely, does an AI Manager's authority improve the performance when the workers are themselves LLM agents?

The direct evidence on LLM agent teams is contradictory. Muralidharan et al.\ \cite{muralidharan2025lessons} report that flat teams outperform hierarchical ones on four commonsense and social-reasoning tasks. OrgAgent \cite{wang2026orgagent}, which layers a system into governance, execution, and compliance tiers, reports the opposite on reading comprehension and multi-hop question answering. Zhang et al.\ \cite{zhang2024socialpsych} find that the better organizational form depends on agent traits, and the largest study of multi-agent failures traces most breakdowns to specification problems and inter-agent misalignment rather than to insufficient supervision \cite{cemri2025failures}. We try to extend this literiture by improving on two features of this literature keep. First, comparisons vary an entire framework rather than one link, so the effect of hierarchy is confounded with prompts, roles, and tooling. Second, they are scored on benchmarks with verifiable short answers, where a Manager can check a candidate answer cheaply. In most use cases, people deploy Manager agents on open-ended synthesis, where quality is multidimensional and no answer key exists.

We isolate the authority link on such a task. Two organizational forms share the same five roles (Researcher, Analyst, Writer, Critic, Manager), the same prompts, tools, model pool, and data. They differ in exactly one authority link. In the hierarchical form the Manager holds \emph{loop-back authority}, the power to reject a worker's output and oblige that worker to revise, capped at two loops per run; in the flat form the Manager is a peer that routes work forward and may only comment. The task is business-intelligence reporting from product specifications and customer reviews, a knowledge-synthesis task with a partially verifiable ground truth: the specifications a report asserts can be checked deterministically, its strategic content cannot. Each of 43 products is run once under each form with an identical random assignment of models to roles, and every report is scored by a five-model LLM judge panel, which rates Writing Clarity and Utility, and by a specification-extraction script.

The flat form scores higher on Utility ($d = 0.42$, $p = 0.009$, robust to dropping any judge) and on Writing Clarity ($d = 0.34$, $p = 0.030$), and the hierarchical form costs 51.5\% more tokens. The classical prediction, that the authority link raises Writing Clarity, fails; the Utility result is the one both literatures expected. The mechanism sits inside the revision loop: reports from the two forms are the same length, but hierarchical reports use 53\% more hedging words such as ``may'' and ``could'', each Manager-issued loop is associated with a 0.14-point drop in Writing Clarity, and the hierarchical Writer's first draft is statistically indistinguishable from the flat report on every text measure. On the one component with a checkable ground truth, specification accuracy, both forms are at ceiling.

The paper makes two contributions. First, it provides a precise test of authority in the context of multi-agent systems through a paired experimental design that isolates a single authority link in an LLM agent team while holding roles, prompts, tools, models, and data fixed, and that pairs a five-model judge panel with a deterministic specification check. The findings shed light on the role of authority and the design of such levers in multi-agent teams. Second, it provides evidence that loop-back authority costs 51.5\% more tokens without improving any quality metric, and that the loss in quality tracks the number of revision loops rather than the presence of the authority link alone.


\section{Related Work}

\subsection{Organizational Design in Multi-Agent Systems}

Organizational design has been a first-class variable in MAS since the Contract Net Protocol allocated tasks by negotiation rather than command \cite{smith1980contractnet}. Horling and Lesser \cite{horling2004survey} survey the major paradigms (hierarchies, holarchies, coalitions, teams, congregations, federations) and show that the choice has quantitative consequences for performance, robustness, and communication cost. Carley and Gasser \cite{carley1999cot} set out the computational organization theory programme. Organizational self-design lets systems restructure as load and environment shift \cite{ishida1992osd}, and So and Durfee \cite{so1996designing} analyse when tree-structured organizations of computational agents pay for themselves. Organizational specification languages such as MOISE+ \cite{hubner2002moise} make the design explicit. An organization is described by a structural specification, which lists roles, groups, and the links between them (acquaintance, communication, or authority), and by a deontic specification, which states what each role is obliged or permitted to do. We adopt this vocabulary for our own design. Our two forms share every role and every communication link. They differ in one authority link: in the flat form the Manager is only permitted to comment, while in the hierarchical form its directive is an obligation on the worker. This tradition established that the right organization depends on the environment. It did not have to ask whether the agent's response to authority is itself a variable, which is the question that generative agents raise.

\subsection{Coordination and Influence Structures}

Malone and Crowston \cite{malone1994coordination} define coordination as managing dependencies between activities, and Simon \cite{simon1962architecture} supplies the near-decomposability logic behind splitting a synthesis task into research, analysis, writing, and critique. Kocak, Levinthal, and Puranam \cite{kocak2022dual} add a dynamic theory of how influence structures shape the dual challenge of search and coordination. In their model two properties of a team's beliefs matter: how similar the agents' beliefs become, and how sharp and decisive they are, which they call resolution. Their flat team (dense, symmetric influence) achieves high similarity but low resolution, converging on vague, averaged positions. Their hierarchical team, anchored by a leader's stable belief, achieves both similarity and resolution, but it risks what they call entrapment: it converges so fast on a suboptimal answer that search stops. This is the exploration-exploitation trade-off \cite{march1991exploration}, seen at the level of team structure. We preserve the framework's independent variable, influence asymmetry, defined as whether one agent can reject and force revision of another's output. In our design this is the authority link. Two assumptions of their framework do not transfer. First, its flat team is a simultaneous, fully connected network, whereas LLM agents generate sequentially and read an accumulated shared state. Our flat form therefore represents the absence of loop-back authority, not symmetric broadcast. Second, its agents respond to authority with genuine belief revision, whereas the LLM literature below suggests that LLM agents may respond with compliance.

\subsection{LLM Agent Teams}

Frameworks such as MetaGPT \cite{hong2023metagpt}, AutoGen \cite{wu2023autogen}, and AgentOrchestra \cite{zhang2025agentorchestra} established that specialised agent teams improve output on complex tasks, and surveys treat hierarchical orchestration under a Manager as the default \cite{tran2025multiagent, guo2024survey}. Tran et al.\ \cite{tran2025multiagent} systematize collaboration across four paradigms (memory, report, relay, debate) and four topologies (bus, star, ring, tree), and topology learning optimises the communication graph directly \cite{zhuge2024gptswarm, yang2025topology}. This work varies whole systems or learns graphs end to end. It does not hold the agents fixed and vary a single authority link, which is the manipulation the organizational tradition asks for.

A second line of work concerns how LLMs respond to critique. RLHF-trained models tend to comply when corrected by an authoritative prompt, a behaviour known as sycophancy \cite{sharma2024sycophancy, wei2023simple, bai2022constitutional}, and under iterative feedback they lose the capacity for divergent thought and retreat to safe, hedged text \cite{liang2024encouraging}. In an iterative generate-and-select loop, additional rounds alone do not raise creativity scores, but the evaluator inside the loop does \cite{anderson2026recipes}. In multi-agent debate, extended interaction produces problem drift far more often on generative tasks than on reasoning tasks with checkable answers \cite{becker2026drift}. Self-correction designs address this by separating generation from critique and grounding critique in tools \cite{gou2024critic, shinn2023reflexion}. Our Critic follows that pattern in both forms, so the manipulation is the authority link, not the presence of critique.

\subsection{Competing Predictions}

The two literatures make opposite predictions for the same manipulation. The classical prediction is that granting the Manager loop-back authority raises Writing Clarity: directive critique forces workers to develop clearer positions, and iterative revision polishes the report \cite{kocak2022dual, boussioux2025socratic}. The LLM-specific prediction is that the same authority lowers Writing Clarity, because coercive revision under an authoritative prompt induces sycophantic compliance and Degeneration-of-Thought, producing hedged and padded text rather than sharpened reasoning \cite{liang2024encouraging, sharma2024sycophancy}. For Utility the two agree in direction: the flat form, without the pressure of forced convergence, should preserve exploratory reasoning and score higher on strategic depth, while the hierarchical form risks entrapment \cite{kocak2022dual}. Our paired experiment tests both predictions.


\section{System Design and Experimental Setup}

\subsection{Agents, Roles, and Communication Topology}

\begin{figure}[t]
\centering
\begin{tikzpicture}[
  font=\scriptsize,
  role/.style={draw, rounded corners=1.5pt, minimum width=1.3cm, minimum height=0.44cm, fill=white, inner sep=1.5pt},
  mgr/.style={draw, rounded corners=1.5pt, fill=gray!15, inner sep=2pt, align=center},
  bb/.style={draw, fill=gray!8, minimum width=6.7cm, minimum height=0.36cm, inner sep=1.5pt},
  fwd/.style={-{Latex[length=1.5mm]}, semithick},
  route/.style={-{Latex[length=1.2mm]}, gray!65, thin},
  loop/.style={-{Latex[length=1.7mm]}, red!70!black, very thick, dashed},
  lab/.style={font=\scriptsize, inner sep=1pt, text width=7.5cm, align=left}]
\node[role] (r1) at (0,0) {Researcher};
\node[role] (a1) at (1.75,0) {Analyst};
\node[role] (w1) at (3.5,0) {Writer};
\node[role] (c1) at (5.25,0) {Critic};
\node (e1) at (6.45,0) {end};
\node[mgr] (m1) at (2.625,1.15) {Manager as peer router\\ reads the latest output, comments only};
\node[bb] (b1) at (2.625,-0.72) {Shared blackboard: read the state, append output};
\draw[fwd] (r1) -- (a1); \draw[fwd] (a1) -- (w1); \draw[fwd] (w1) -- (c1); \draw[fwd] (c1) -- (e1);
\foreach \n in {r1,a1,w1,c1} \draw[route] (m1) -- (\n);
\node[lab, anchor=north west] at (-0.7,-1.02) {(a) Flat form: forward routing only. The Manager's prompt states that it has no authority to reject work or force any worker to redo a task.};
\begin{scope}[yshift=-3.45cm]
\node[role] (r2) at (0,0) {Researcher};
\node[role] (a2) at (1.75,0) {Analyst};
\node[role] (w2) at (3.5,0) {Writer};
\node[role] (c2) at (5.25,0) {Critic};
\node (e2) at (6.45,0) {end};
\node[mgr] (m2) at (2.625,1.15) {Manager with loop-back authority\\ reads the entire blackboard, directives bind};
\node[bb] (b2) at (2.625,-0.72) {Shared blackboard: read the state, append output};
\draw[fwd] (r2) -- (a2); \draw[fwd] (a2) -- (w2); \draw[fwd] (w2) -- (c2); \draw[fwd] (c2) -- (e2);
\foreach \n in {r2,a2,w2,c2} \draw[route] (m2) -- (\n);
\draw[loop] (m2.west) to[bend right=35] (r2.north);
\draw[loop] (m2.south west) to[bend right=25] (a2.north west);
\draw[loop] (m2.south east) to[bend left=25] (w2.north east);
\node[lab, anchor=north west] at (-0.7,-1.02) {(b) Hierarchical form: after any step the Manager may reject the output and oblige that role to revise (dashed; at most two loops per run). The Manager ends the run, at once if the Critic approves.};
\end{scope}
\end{tikzpicture}
\caption{The two organizational forms. Roles, prompts, tools, models, data, and the blackboard are identical; the forms differ in one authority link between the Manager and the workers (dashed red).}
\Description{Two-panel schematic. Each panel shows four worker roles in a forward chain (Researcher, Analyst, Writer, Critic) above a shared blackboard, with a Manager node above the chain connected to every worker by thin routing arrows. In the hierarchical panel, additional dashed red arrows run from the Manager back to the Researcher, Analyst, and Writer, representing loop-back directives.}
\label{fig:system}
\end{figure}
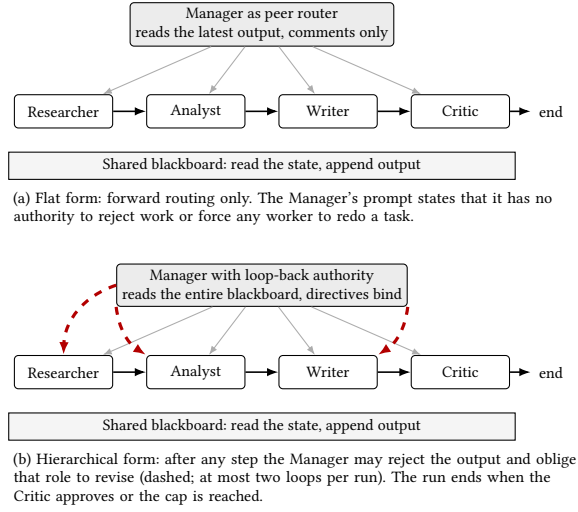

Both organizational forms are implemented in LangGraph \cite{langgraph2024} as state machines over a shared blackboard, the classical MAS coordination artefact through which every agent reads the accumulated state and appends its own output (Figure~\ref{fig:system}). Four worker roles are identical across forms. The Researcher retrieves product metadata and customer reviews with two tools, \texttt{get\_product\_specs} and \texttt{search\_reviews}, the latter a semantic search with sentiment filtering over a ChromaDB store \cite{chroma2024} with Sentence-BERT embeddings \cite{reimers2019sentencebert}, a retrieval-augmented design \cite{lewis2020rag}. The Analyst identifies patterns and root causes without tools. The Writer produces the report, with a required Technical Specifications section. The Critic reviews the draft and calls \texttt{verify\_claim}, which scores each checked assertion against the vector store by embedding similarity and labels it VERIFIED, PARTIAL, or UNVERIFIED (thresholds 0.7 and 0.5); the labels inform the Critic and Manager and enter no metric. Separating generation from tool-grounded critique follows CRITIC and Reflexion \cite{gou2024critic, shinn2023reflexion} and prevents a single model from confirming its own hallucinations; it is present in both forms. Each worker is a ReAct agent \cite{yao2023react}. The fifth role, the Manager, is the routing node; its coordination policy is the treatment (Section 3.2).

To avoid tying results to one model family, each role is filled from a pool of five frontier models (GPT-5.4, Gemini-3.1-Pro, Qwen-3.5-122B, GLM-5, Mistral-Large-3) by random assignment; the Manager draws from the four-model subset without Qwen, whose thinking-mode output stream is incompatible with the structured-output schema used for routing. Temperature is 0, API endpoint versions are pinned, and the model identifier returned is recorded. For each product the same role-to-model assignment is used in both forms, so model identity is a within-pair constant. Implementation details are in supplementary material Section G.

\subsection{The Treatment: Loop-Back Authority}

\begin{table}[t]
\caption{The one link that differs between the forms, and everything that does not.}
\label{tab:treatment}
\centering\footnotesize
\setlength{\tabcolsep}{3pt}
\begin{tabular}{@{} >{\raggedright\arraybackslash}p{1.75cm} >{\raggedright\arraybackslash}p{2.95cm} >{\raggedright\arraybackslash}p{3.15cm} @{}}
\toprule
 & \textbf{Flat} & \textbf{Hierarchical} \\
\midrule
Manager reviews & the latest worker output & the entire blackboard after every step \\
\addlinespace[2pt]
Manager output & non-binding commentary (a permission) & a directive naming what must be fixed, injected into the worker's context (an obligation) \\
\addlinespace[2pt]
Prompt clause & ``You do NOT have authority to reject work or force any worker to redo their task'' & ``You have absolute authority to force a Loop-Back if the current output is corrupted'' \\
\addlinespace[2pt]
Revision loops & none (counter fixed at 0) & Manager-chosen, at most 2 per run \\
\addlinespace[2pt]
Run ends & after the Critic & on Critic approval, or when the Manager ends it \\
\addlinespace[2pt]
Maximum steps & 9 & 15 \\
\addlinespace[2pt]
Worker roles, prompts, tools, models, data & identical & identical \\
\bottomrule
\end{tabular}
\end{table}

The two forms differ in one authority link and its deontic reading (Table~\ref{tab:treatment}). In the flat form the Manager is a peer dispatcher: it enforces the forward sequence Researcher, Analyst, Writer, Critic, end; reviews only the latest worker output; and may add non-binding commentary. Its prompt states that it does not have authority to reject work or force any worker to redo a task, and the loop counter is fixed at zero. In the hierarchical form the Manager reviews the entire blackboard after each step, applies an explicit decision guide (after the Researcher, for example, ``Does the data include product specs AND customer reviews with quotes?''), and either proceeds or issues a loop-back directive that names what must be fixed and is injected into the target worker's context. A directive is an obligation on the worker, not a suggestion. The Manager may loop back at most twice per run. A Critic APPROVED flag ends the run at once; otherwise the Manager decides when to end it, and in 13 of the 43 runs it did so below the cap without an approval. Because the same worker prompts serve both forms, the manipulation is a change of interaction protocol between roles, not a change of any role. The authority link carries no sanction: a worker that ignored a directive would suffer nothing, so any compliance we observe is a property of the model's response to authoritative instruction rather than of enforced incentives. The full Manager prompt is in supplementary material Section F.

\subsection{Task and Data}

The task is business-intelligence reporting about e-commerce products: a multi-agent knowledge-synthesis task whose ground truth is partially verifiable, since a report's technical specifications can be checked against the listing while its analytical content cannot. Automated synthesis of unstructured customer feedback is operationally common \cite{liu2012sentiment}, and the task carries both halves of the dual challenge, search (finding the relevant facts) and coordination (aligning them into a coherent narrative) \cite{kocak2022dual}. We collected gaming laptops with the RTX 4060 GPU from the Amazon US marketplace; a single GPU class controls hardware heterogeneity. After de-duplication and removal of short reviews the pool held 52 products with 540 reviews, and the 43 products with at least five reviews entered the paired analysis. Reviews are stratified by star rating (4 or above positive, 3 or below negative) so that retrieval is not skewed positive. Each product is run once under each form under identical conditions, giving 86 runs; an efficiency tracker records tokens, API cost, and latency. Data provenance and the ethics statement are in the supplementary material.

\subsection{Evaluation and Metrics}

Subjective quality is scored by a five-model judge panel (the same five models, each judging every report), following the LLM-as-a-judge paradigm \cite{liu2023geval, zheng2023judging} with two mitigations for its known biases: a panel rather than a single judge, and a qualitative-first protocol in which each judge writes an analysis before assigning a 1 to 5 score \cite{choudhury2025wadetest}. Generative models have been shown to evaluate strategic decisions reliably \cite{doshi2024genai}. Writing Clarity averages Structure, Coherence, and Conciseness; Utility averages Actionability, Root Cause Analysis, and Strategic Depth (rubric anchors in supplementary material Section E). We read Writing Clarity as the exploitation-side outcome and Utility, through Strategic Depth, as the exploration-side outcome; this mapping is an interpretive assumption, not a validated correspondence, and we report every dimension. Specification accuracy is scored deterministically, in the spirit of FActScore \cite{min2023factscore} but restricted to checkable content: regex patterns extract the specifications a report asserts (GPU, RAM, CPU, storage) and verify each against the listing metadata, scaled to 1 to 5. This original metric scores asserted specifications only and never penalizes omission; Section 4.3 also reports a repaired version that scores precision and recall over all listing specifications with exact variant matching. Final Score is the mean of the three pillars.

\subsection{Analysis}

The primary tests are paired-sample $t$-tests on Writing Clarity and Utility across the 43 product pairs, with the paired-sample Cohen's $d$ (mean of differences over their SD). A linear mixed-effects model on all 86 runs, with form, writer model, number of reviews, and mean rating as fixed effects and product as a random intercept, checks that the paired effects are not artefacts of model assignment or data availability. To locate the mechanism we fit, within the hierarchical arm, an OLS of quality on the number of Manager-issued loops with the same controls; compare the paired difference in the subset that finished with zero loops; and compare hierarchical Writer drafts with each other and with the flat report on deterministic text measures (hedging density, lexical diversity, and overlap between successive drafts). Judge reliability is reported as Krippendorff's alpha \cite{krippendorff2011alpha}, per-judge directional concordance, and leave-one-judge-out re-estimation. With 43 pairs the design has 80\% power for $d = 0.44$; per-dimension follow-ups are exploratory. Wilcoxon and Holm-corrected \cite{holm1979simple} versions of every primary test are in the supplement.


\section{Results}

\subsection{Overall Comparison}

\begin{table}[t]
\caption{Paired comparison across the 43 product pairs (top) and the mixed-effects estimate of the form effect on all 86 runs (bottom). $d$ is the paired-sample Cohen's $d$, positive when Flat scores higher. The mixed-effects model controls for writer model, number of reviews, and mean rating, with product as a random intercept.}
\label{tab:main}
\centering\footnotesize
\setlength{\tabcolsep}{3.5pt}
\begin{tabular}{@{} l c c c c @{}}
\toprule
Metric & Flat $M$ (SD) & Hier.\ $M$ (SD) & $d$ & $p$ \\
\midrule
Final Score      & 4.563 (0.233) & 4.508 (0.217) & 0.262    & 0.097 \\
Writing Clarity  & 4.454 (0.137) & 4.360 (0.238) & 0.342    & 0.030 \\
Utility          & 4.715 (0.148) & 4.621 (0.208) & 0.417    & 0.009 \\
Spec.\ accuracy  & 4.520 (0.615) & 4.543 (0.626) & $-$0.046 & 0.764 \\
\bottomrule
\end{tabular}

\smallskip
\begin{tabular}{@{} l c c c @{}}
\toprule
Mixed effects, Hierarchical $= 1$ & $\beta$ & SE & $p$ \\
\midrule
Writing Clarity  & $-$0.094 & 0.042 & 0.025 \\
Utility          & $-$0.094 & 0.034 & 0.006 \\
Spec.\ accuracy  & $+$0.023 & 0.076 & 0.763 \\
\bottomrule
\end{tabular}
\end{table}

All 86 runs completed, in about 5.9 hours at an estimated cost of \$47.71. Table~\ref{tab:main} gives the paired comparison and the mixed-effects estimate of the form effect. The flat form scores higher on Writing Clarity ($d = 0.34$, $p = 0.030$) and on Utility ($d = 0.42$, $p = 0.009$), wins on Final Score in 28 of 43 products (hierarchical 14, one tie), and does not differ on specification accuracy. Both primary tests survive Holm correction (supplementary material Section A); under the Wilcoxon signed-rank alternative the Utility test survives and the Writing Clarity test is borderline ($p = 0.051$). Once writer model, number of reviews, and mean rating are partialled out and product is a random intercept, the hierarchical form is associated with lower Writing Clarity ($\beta = -0.094$, $p = 0.025$) and lower Utility ($\beta = -0.094$, $p = 0.006$), and with no change in accuracy. The classical prediction, that the authority link raises Writing Clarity, is rejected in the direction the LLM-specific prediction anticipated; the shared expectation for Utility is confirmed.

\subsection{Where the Gap Comes From}

\begin{figure}[t]
\centering
\includegraphics[width=\columnwidth]{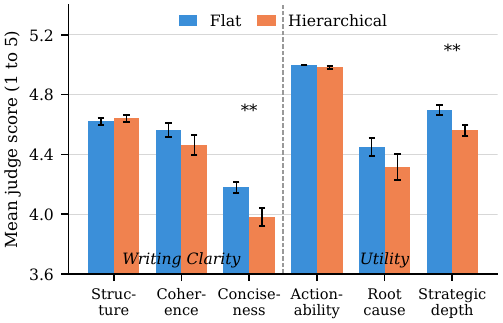}
\caption{Per-dimension mean judge scores by form, with standard-error bars ($N = 43$ pairs). Two asterisks mark $p < 0.01$ in the paired test.}
\Description{Grouped bar chart of Flat and Hierarchical mean scores on Structure, Coherence, Conciseness, Actionability, Root Cause Analysis, and Strategic Depth, with standard-error bars; Conciseness and Strategic Depth are marked as significant.}
\label{fig:dimensions}
\end{figure}

Figure~\ref{fig:dimensions} decomposes both composites. Within Writing Clarity the gap is carried entirely by Conciseness (flat 4.18 versus hierarchical 3.98, $d = 0.44$, $p = 0.007$); Structure ($d = -0.09$, $p = 0.56$) and Coherence ($d = 0.19$, $p = 0.22$) do not differ. The Structure null is not evidence of equivalence: a two one-sided test rejects only at a lenient bound ($|d| < 0.35$, $p = 0.048$), not at the conventional $|d| < 0.20$ ($p = 0.24$), and 84 to 88\% of reports score 4.5 or above on Structure, a ceiling (supplementary material Section H). What the judges penalised is therefore not disorganised writing but padded writing, and it is not a length effect. Flat and hierarchical reports contain the same number of words (1036 versus 1046 on average, paired $p = 0.76$); the token overhead of the hierarchical form is pipeline traffic, not report length. Regressing each product's paired score difference on its paired word-count difference leaves every effect where it was (Writing Clarity at equal length $+0.090$, $p = 0.031$; Utility $+0.096$, $p = 0.008$; Conciseness $+0.193$ against $+0.197$ raw; supplementary material Section I). What differs is stance. Measured with a fixed lexicon of epistemic hedges, hierarchical reports carry 53\% more hedges per 1000 words than flat reports on the same product (5.03 versus 3.30, $d_z = 0.61$, $p < 0.001$), spread across the modal and evidential core of the lexicon (\emph{possible}, \emph{may}, \emph{suggests}, \emph{could}) rather than produced by one term, while lexical diversity is identical between forms (MTLD 186 versus 189, $p = 0.65$). Section 4.5 shows where in the process this hedging is introduced.

Within Utility, Strategic Depth carries the effect (4.70 versus 4.56, $d = 0.48$, $p = 0.003$) and is the only per-dimension test that survives Holm correction across the six follow-ups (Holm $p = 0.021$); Root Cause Analysis leans the same way ($d = 0.23$, $p = 0.14$) and Actionability is at ceiling in both forms (98\% of scores at or above 4.9). Read through the framework, this is the pattern the shared Utility prediction anticipates: the form without forced convergence produces deeper strategic content, and the form with it converges on shallower content, which is Kocak et al.'s entrapment on a non-peak. Because the rubric was built for this study and Strategic Depth is a single dimension from a single dataset, we treat the size of the effect, not its existence, as provisional.

\subsection{Specification Accuracy: No Difference, and a Ceiling}

Specification accuracy does not differ between forms on the original metric ($d = -0.05$, $p = 0.76$; 60 to 63\% of reports have every asserted specification verified in both forms), and the mixed-effects model attributes its variance to writer model (Mistral-Large-3 $\beta = +0.65$, $p = 0.020$) and to the number of available reviews ($p = 0.068$), not to structure. The original metric, however, understates how little there is to explain. It scores only the specifications a report happens to assert, matches loosely, and imputes the scale midpoint when nothing checkable is asserted. We rebuilt it against a validated per-product ground truth (five to seven specifications per listing: RAM, storage, GPU, CPU, display size, refresh rate, price), requiring exact hardware variants and scoring both precision and recall over all listing specifications (supplementary material Section K). On the repaired metric both forms are at ceiling: mean recall and precision are 1.000 in the flat form and 0.997 in the hierarchical form, 85 of 86 reports restate every listing specification correctly, and the single error in the corpus is one hierarchical report that names a CPU model absent from the listing. The variance of the original metric was parser noise (GPU memory matched as system RAM, comparison mentions read as claims), which is why it correlates only $r = 0.15$ with the repaired score.

The null is therefore not a failure to detect an effect; it is a component of the task on which there was nothing for a supervisor to fix. That matters for reading the disagreement in the literature. Reports of hierarchical advantage come from benchmarks with verifiable short answers, where a compliance or verification tier can cheaply check a candidate answer and correct abstention is itself scored \cite{wang2026orgagent}; reports of flat advantage come from open-ended and social-reasoning settings \cite{muralidharan2025lessons}; and multi-agent debate drifts far more on generative than on checkable tasks \cite{becker2026drift}. Our results fall on that seam. On the checkable component, structure makes no difference; on the open-ended components, the authority link is associated with harm.

\subsection{Cost}

The hierarchical form consumed 51.5\% more tokens per report (74,781 against 49,370), had an estimated 40.5\% higher generation cost and 20.2\% higher total cost (generation plus evaluation, \$0.606 against \$0.504), and ran 34.3\% slower (282 against 210 seconds), with high variance on every measure because loops are Manager-assigned (full table in supplementary material Section H). Against this overhead it delivered no gain on any of the three quality metrics. In this setting the authority link is pure cost.

\subsection{Is the Harm from Hierarchy Itself, or from its Revision Loops?}

\begin{figure}[t]
\centering
\includegraphics[width=\columnwidth]{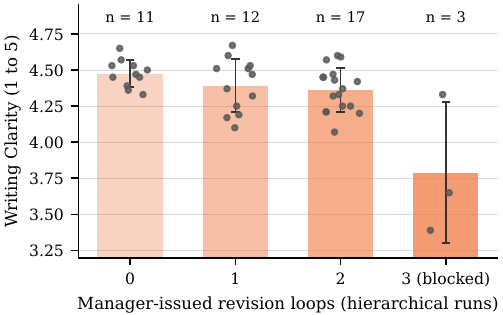}
\caption{Writing Clarity by the number of Manager-issued revision loops in the 43 hierarchical runs: bucket means with standard-deviation bars, individual runs as dots. The three runs at three loops had a third revision request blocked by the cap.}
\Description{Bar chart of mean Writing Clarity for hierarchical runs with zero, one, two, and three revision loops, declining from 4.48 to 3.79, with individual runs overlaid as dots.}
\label{fig:loops}
\end{figure}

The authority link is present in every hierarchical run, but the Manager exercised it in only 32 of 43: 11 runs finished with zero loops, 12 with one, 17 with two, and 3 with a third request that the cap blocked. Loops count Manager re-delegations to any worker; the Writer produced one draft in 12 runs, two in 28, and three in 3. Because loops are zero in every flat run by construction, structure and loops cannot enter one regression, so we use three views.

\emph{Dose-response within the hierarchical arm.} Writing Clarity falls monotonically with loop count (Figure~\ref{fig:loops}; Spearman $\rho = -0.40$, $p = 0.008$). An OLS with writer model, number of reviews, and mean rating as controls estimates $-0.142$ Writing Clarity points per loop ($p < 0.001$, $R^2 = 0.41$); for Utility the slope is negative but not significant ($\beta = -0.050$, $p = 0.18$), and for accuracy it is null. The coefficient is correlational: the Manager chooses which drafts to send back, so weak first drafts attract loops. It also leans on the three cap-blocked runs; dropping them halves the slope ($\beta = -0.074$, $p = 0.038$) and recoding them to two loops gives $-0.132$ ($p = 0.006$). In a specification that adds report length as a covariate the per-loop association remains ($\beta = -0.10$, $p = 0.008$), so revision costs more than concision (supplementary material Sections H and I).

\emph{The zero-loop subset.} In the 11 pairs where the Manager never intervened, the hierarchical report is numerically higher than the flat one on all three metrics (Writing Clarity $d = -0.30$, Utility $d = -0.42$, neither significant at $n = 11$). An authority link that was never exercised did no measurable harm. These products were selected by the Manager's own judgement, however, so the subset triangulates rather than tests.

\emph{Inside the loop: successive drafts.} The cleanest evidence holds product, models, and roles fixed and varies only the act of revision, by extracting every Writer draft from the hierarchical run logs (77 drafts; the final draft matches the stored report in 43 of 43 runs). Before any Manager intervention, the hierarchical Writer's first draft is statistically indistinguishable from the flat report for the same product on every measure: hedging density $d_z = 0.11$ ($p = 0.46$), lexical diversity $p = 0.98$, length $p = 0.20$. The entire hedging gap between the forms therefore opens after the first draft. Within the 31 runs that revised, the second draft hedges 1.95 more per 1000 words than the first ($d_z = 0.83$, $p < 0.001$) and is 38 words longer ($p = 0.007$), while lexical diversity barely moves (MTLD $p = 0.26$). And revision is not rewriting: the median revised draft keeps 88\% of its predecessor as a common subsequence and 85\% of its content vocabulary, none of the 34 draft transitions is a substantial rewrite (the least conservative keeps 51\%), and one returned the draft unchanged (supplementary material Section L).

Together the three views locate the harm. The mixed-effects estimate attributes a small, stable penalty to the hierarchical form ($\beta \approx -0.09$ on both composites); the dose-response and zero-loop views attribute that penalty to the runs in which the authority link was exercised; and the draft-level comparison shows what exercising it does to the text. Told to revise, the Writer does not reconsider the analysis. It keeps the draft, softens its claims, and appends qualifying material, and the judges penalise the result as padding. This is a direct fit to sycophantic compliance \cite{sharma2024sycophancy}. It is a poorer fit to Degeneration-of-Thought in its strong form, which implies progressive drift into worse content \cite{liang2024encouraging}: there is no drift, no vocabulary collapse, and no rewriting. The loop does not make the agent think worse; it makes the agent commit less.

\subsection{Judge Reliability and Robustness}

\begin{table}[t]
\caption{Robustness of the two headline results. LOJO: leave-one-judge-out re-estimation on the four remaining judges (range over the five omissions). Equal-length $\alpha$: the paired effect at zero word-count difference.}
\label{tab:robust}
\centering\footnotesize
\setlength{\tabcolsep}{4pt}
\begin{tabular}{@{} l c c @{}}
\toprule
Check & Writing Clarity & Utility \\
\midrule
Paired $t$, $p$ (Holm)          & 0.030 (0.030)    & 0.009 (0.018) \\
Wilcoxon, $p$                    & 0.051            & 0.009 \\
Mixed effects, $\beta$ ($p$)     & $-$0.094 (0.025) & $-$0.094 (0.006) \\
Equal-length $\alpha$ ($p$)      & $+$0.090 (0.031) & $+$0.096 (0.008) \\
Judges individually favouring Flat & 4 of 5         & 5 of 5 \\
LOJO $d$, range                  & 0.27 to 0.40     & 0.38 to 0.43 \\
LOJO $p$, range                  & 0.012 to 0.083   & 0.007 to 0.016 \\
\bottomrule
\end{tabular}
\end{table}

Absolute agreement among the five judges is low, Krippendorff's alpha 0.08 to 0.17 for Writing Clarity and 0.22 to 0.30 for Utility, driven mainly by GPT-5.4's severity (mean 3.86 against a panel range of 4.44 to 4.68; excluding it raises alpha to 0.26 and 0.34). The paired design does not need calibration agreement, it needs directional agreement, and that is present: every judge agrees with the panel-mean direction on at least 68\% of pairs for Writing Clarity and 77\% for Utility, three of five exceed 80\% on Utility, and the paired-difference correlations between judges are uniformly positive (supplementary material Sections B and D).

Leave-one-judge-out re-estimation separates the two headline results (Table~\ref{tab:robust}; full tables in supplementary material Section J). Utility is robust: all five judges individually favour the flat form ($d$ from 0.25 to 0.32), the four-judge panel gives $d$ between 0.38 and 0.43 with $p \leq 0.016$ whichever judge is dropped, and Strategic Depth stays at $p \leq 0.010$ under every omission. Writing Clarity is judge-sensitive: it is carried by Qwen, GLM, and Gemini, GPT-5.4 individually leans the other way ($d = -0.14$, $p = 0.38$), and dropping any of the three carriers moves the composite to $p$ between 0.055 and 0.083, consistent with its Wilcoxon $p = 0.051$. Conciseness, the dimension behind the composite, remains at $p \leq 0.021$ under every omission. We therefore lead with Utility and Strategic Depth as the robust result and report the Writing Clarity effect as real in direction and fragile in significance. The mechanism evidence in Section 4.5 rests on deterministic text measures and does not depend on the judges at all.

Two human raters, blind to condition, compared the flat and hierarchical reports for 20 randomly drawn products, choosing one report (or a tie) on overall preference, Conciseness, and Strategic Depth. Pooled over both raters, the flat report won 54\% of non-tie overall choices (21 of 39, binomial $p = 0.75$), 57\% for Strategic Depth and 55\% for Conciseness. The direction matches the panel on all three questions, but 20 pairs are too small for a statistically robust inference and cannot confirm it. Further, the raters split: one preferred the flat report in 14 of 20 pairs, the other in 7 of 19. The two raters agreed modestly with each other (Cohen's $\kappa = 0.30$ for overall preference, 0.28 for Strategic Depth, $-0.18$ for Conciseness), within the range of agreement between pairs of judges on the same products ($\kappa$ from 0.00 to 0.38 on overall preference). Low agreement is therefore partly a property of the task, i.e., these reports are hard to rank for people as well as for models.


\section{Discussion and Conclusion}

We asked whether the authority link that production frameworks install by default, a Manager that can reject work and oblige revision, helps a team of LLM agents on open-ended synthesis. Holding everything else fixed across 43 paired products, it does not. The flat form scores higher on Utility and on Writing Clarity, the hierarchical form costs half again as many tokens, and on the one checkable component, specification accuracy, both are at ceiling. The harm is not a property of the form as such: the hierarchical Writer's first draft is indistinguishable from the flat report, and runs in which the link is never exercised show no measurable deficit. It is a property of what the link does when exercised. Directed to revise, the Writer complies by hedging and padding while keeping the draft intact, and the judges penalise the result.

For multi-agent system design the finding gives the tradition's contingency claim \cite{horling2004survey} a mechanism specific to generative agents. In a classical organization an authority link is a control channel; for an RLHF-trained worker it is also a conditioning signal, and an unverifiable critique is treated as a social cue to comply with rather than as information to act on. Since the link here carried no sanction, the compliance we observe is the model's own. The design rule that follows is that a Manager with loop-back authority earns its cost when it can verify something and becomes a liability when it can only opine. Concretely, for short-context synthesis a flat sequential pipeline should be the default; if a Manager is needed for decomposition and routing, it should not hold loop-back authority; and revision should be triggered by a verifier flag (a failed \texttt{verify\_claim}, a missing section) rather than by managerial judgement of quality. Whether the factuality of analytical claims, which we did not measure, is equally insensitive to the form is open.

For coordination theory the transfer is asymmetric. The shared Utility prediction holds: the flat form scores higher on strategic depth, as entrapment predicts \cite{kocak2022dual}. The classical prediction on Writing Clarity does not: loop-back authority brings no gain in structure or coherence and a loss in concision, as the LLM-specific prediction expected. The divergence sits in an assumption the theory builds in rather than varies, that workers respond to authority with belief revision. Human writers demonstrably adapt to the audiences that evaluate them \cite{zohrehvand2022feedback}; language models adapt by softening. The contrast with the adjacent cell of the design space, where an AI manager improves a team of human workers \cite{zohrehvand2026coordination}, suggests the failure is specific to a loop in which both the Manager that demands the revision and the Writer that performs it are RLHF-trained models.

\emph{Limitations.} The regime is one short-context synthesis task in one product domain, one generation of five frontier models, one Manager prompt, and 43 pairs; we do not extend the design rule to long-context generation, code, or tasks with richer verifiers. Quality is scored by LLM judges that share RLHF-style preferences and may jointly penalise a style humans would not. A blinded check with two human raters on 20 pairs leaned the same way as the judges but was too small to confirm the direction, and the two raters disagreed with each other. The paired design cancels judge effects that do not interact with the treatment, the Utility result survives every leave-one-judge-out check, and the mechanism evidence is judge-free, but the Writing Clarity composite is judge-sensitive and should be read as such. The loop-count and zero-loop analyses are correlational because the Manager assigns its own loops; the draft-level comparison is the estimate to lead with. The hedge lexicon cannot distinguish calibrated uncertainty from evasion, so the reading of hedging as harm rests on the quality scores.

Three follow-ups are immediate. Making the revision trigger the manipulated variable (verifier-fired against Manager-judged loops, single-pass review, peer negotiation without authority) would test the design rule; introducing a sanction or a persistent stake for the worker would test whether compliance without deference is the boundary condition that the theory's translation needs; and a measure of deference read from the Writer model's activations rather than from surface words, which concept probes on frozen models now permit \cite{hazenoot2026measuring}, would separate calibrated uncertainty from evasion without relying on the judges.

\emph{Ethics and AI use.} The product pages and reviews are public; no account data was collected, reviewer names are not retained, and the raw scrape is not redistributed. AI assistants were used to debug code, draft, and edit prose; the authors take responsibility for the content. The full statement and the code and data availability are in the supplementary material.

\clearpage


\section*{Supplementary material}

# Supplementary Material for: Loop-Back Authority in LLM Agent Teams: A Paired Experiment on Flat and Hierarchical Coordination

BURAK AGACHAN, Leiden Institute of Advanced Computer Science (LIACS), Leiden University, The Netherlands
MAX VAN DUIJN, Leiden Institute of Advanced Computer Science (LIACS), Leiden University, The Netherlands
AMIRHOSSEIN ZOHREHVAND, Leiden Institute of Advanced Computer Science (LIACS), Leiden University, The Netherlands

This document collects the material referenced from the main paper: the data and ethics statement, the robustness checks, the per-judge concordance figures, the mixed-effects regression output, the paired-difference inter-judge correlations, the construct-validity evidence and rubric anchors, and the verbatim Hierarchical Manager prompt.



## 1 Data Sources and Ethics

### 1.1 Amazon review and product-specification data

The product specifications and customer reviews used in this study were collected via third-party data-collection services (specifically Apify and a paid commercial browser-extension scraper) during January 2026; we did not use Amazon's Product Advertising API. Collection was restricted to publicly visible product pages and customer reviews. We did not retrieve account information, IP addresses, or any non-public metadata. Review author display names that appear on the public page were not used in any analysis and are not retained in the analytic dataset. Reviews are stored only in the form needed for retrieval-augmented generation: review text, star rating, and product ASIN. We acknowledge that automated collection of Amazon page content is in tension with Amazon's terms of service for commercial purposes; the data was used solely for non-commercial academic research and is not redistributed in raw form.

### 1.2 AI-use disclosure

AI assistants were used to debug the analysis and orchestration code, to edit the prose, and to help search the literature. The authors reviewed every output and take full responsibility for the content. The language models inside the studied system and on the judge panel are part of the experimental design, described in the main paper.

### 1.3 Data and code availability

Code is publicly available at https://github.com/cihatburak/Master_Thesis_Multi_Agents. The repository includes the LangGraph implementation of both architectures, the multi-model judge panel, the experiment driver, and all statistical analyses reported in this paper. Generated artefacts (the Chroma vector store, raw model traces, and evaluation outputs) are reproducible by running the scripts in the order documented in the repository README. The Hierarchical Manager prompt is also reproduced verbatim in supplementary material Section F; the per-dimension judge rubrics are in supplementary material Section E. The 43-product analytic dataset and the per-run evaluation outputs are available

Authors' Contact Information: Burak Agachan, Leiden Institute of Advanced Computer Science (LIACS), Leiden University, Leiden, The Netherlands; Max van Duijn, Leiden Institute of Advanced Computer Science (LIACS), Leiden University, Leiden, The Netherlands; Amirhossein Zohrehvand, Leiden Institute of Advanced Computer Science (LIACS), Leiden University, Leiden, The Netherlands.

from the authors on request and will be deposited in the repository; raw scraped material is not redistributed for the terms-of-service reason noted above.

## A Robustness Checks: Wilcoxon and Holm-Bonferroni Correction

The main text reports paired-sample $t$-tests as the primary tests for Writing Clarity and Utility. As a robustness check we also computed Wilcoxon signed-rank tests and Holm-Bonferroni step-down corrections [32] for both. Tables 1 and 2 show that the substantive conclusions are unchanged: both tests survive Holm correction under the paired-$t$ test; under Wilcoxon, the Utility test survives and the Writing Clarity test is borderline ($p = 0.051$). A specification-extraction accuracy test is included for completeness.

Table 1. Primary Tests with Wilcoxon and Holm-Bonferroni Correction

| **Hypothesis** | **paired-$t$ $p$** | **Holm-$t$ $p$** | **Wilcoxon $p$** | **Holm-Wilcoxon $p$** |
|---|---|---|---|---|
| Utility | 0.009 | 0.018* | 0.009 | 0.018* |
| Writing Clarity | 0.030 | 0.030* | 0.051 | 0.051 |
| Accuracy | 0.764 | n/a | 0.717 | n/a |

* $p < 0.05$. Holm step-down applied across the two primary hypotheses. Specification-extraction accuracy is not corrected (reported as a supplementary observation, not a hypothesis); the label *Accuracy* in this paper always refers to specification-extraction accuracy.

Table 2. Per-Dimension Tests with Wilcoxon and Holm-Bonferroni Correction (6 dimensions)

| **Dimension** | **paired-$t$ $p$** | **Holm-$t$ $p$** | **Wilcoxon $p$** | **Holm-Wilcoxon $p$** |
|---|---|---|---|---|
| Strategic Depth | 0.003 | 0.021* | 0.007 | 0.041* |
| Conciseness | 0.007 | 0.033* | 0.017 | 0.084 |
| Root Cause Analysis | 0.144 | 0.577 | 0.181 | 0.724 |
| Coherence | 0.220 | 0.659 | 0.489 | 0.977 |
| Actionability | 0.224 | 0.659 | 0.381 | 1.000 |
| Structure | 0.557 | 0.659 | 0.645 | 0.977 |

* $p < 0.05$. Holm step-down applied across the six per-dimension follow-up tests (WC $\times$ 3 + UT $\times$ 3).

Strategic Depth survives Holm correction under both base tests, Conciseness survives only under the $t$-test (Wilcoxon Holm $p = 0.084$), and the remaining four dimensions are non-significant under both. The conclusion of the Utility results subsection of the main paper is unchanged: Strategic Depth is the dimension on which Flat most clearly outperforms Hierarchical.

## B Per-Judge Directional Concordance

Krippendorff's alpha measures absolute-level agreement and is sensitive to between-judge calibration offsets. Because our paired design tests within-product differences, we additionally report each judge's agreement with the panel-mean direction across the 43 paired comparisons.

Every judge agrees with the panel-mean direction on at least 68% of paired comparisons for Writing Clarity and 77% for Utility, with three of five judges exceeding 80% on both. The judges therefore disagree more on absolute calibration than on the direction of architectural difference per product, which is the question the paired design is built to answer.

Table 3. Per-Judge Concordance with the Panel-Mean Direction (N = 43 pairs)

| Judge | Writing Clarity | Utility |
|---|---|---|
| GPT-5.4 | 68.2% | 87.5% |
| Gemini-3.1-Pro | 91.7% | 91.7% |
| Qwen-3.5-122B | 82.1% | 80.0% |
| GLM-5 | 71.4% | 89.7% |
| Mistral-Large-3 | 77.8% | 77.1% |

## C Architectural Main-Effect Mixed-Effects Regression

To check that the paired-test results are not artefacts of differential writer-model assignment or product-level data availability, we fit a linear mixed-effects model on the full $N = 86$ sample with product as a random intercept and `Score ~ Architecture + WriterModel + NumReviews + MeanRating` as fixed effects.

Table 4. Architectural Main Effect (Hier vs Flat), N = 86, 43 product groups

| Dependent Variable | $\beta$ (Hier = 1) | SE | $p$ |
|---|---|---|---|
| Writing Clarity | −0.094 | 0.042 | 0.025* |
| Utility | −0.094 | 0.034 | 0.006** |
| Accuracy | +0.023 | 0.076 | 0.763 |

* $p < 0.05$, ** $p < 0.01$. WriterModel reference: Gemini-3.1-Pro. Specification-extraction accuracy is additionally predicted by WriterModel (Mistral-Large-3: $\beta = +0.650$, $p = 0.020$) and number of reviews ($\beta = +0.066$, $p = 0.068$).

Once writer-model identity, number of reviews, and mean rating are partialled out, Hierarchical is associated with significantly lower Writing Clarity ($\beta = -0.094$, $p = 0.025$) and Utility ($\beta = -0.094$, $p = 0.006$), and shows no architectural effect on Accuracy ($\beta = +0.023$, $p = 0.763$). The architectural penalty in this main-effect model is small ($\beta \approx -0.09$) compared with the per-loop coefficient inside the hierarchical arm ($\beta = -0.142$ on WC, reported in the dose-response subsection of the main paper), consistent with the decomposition reported in the main text: the composite Writing Clarity gap is consistent with the iterative-revision mechanism doing most of the work rather than the loop-back authority manipulation alone.

## D Paired-Difference Inter-Judge Correlations

Because our design tests within-product paired differences rather than absolute scores, the most informative inter-judge agreement statistic is the correlation across the 43 paired (Flat − Hierarchical) difference scores, computed separately for each judge. Krippendorff's alpha on absolute scores (the Inter-Rater Reliability table of the main paper) penalises calibration offsets that the paired design already cancels; the paired-difference correlation isolates whether the judges agree on the direction and magnitude of the treatment effect per product, which is the quantity our hypotheses operate on. Table 5 reports the 5×5 Pearson correlation matrices for Writing Clarity and Utility.

The off-diagonal correlations are uniformly positive on both metrics (mean $r = 0.29$ for Writing Clarity and 0.29 for Utility), indicating that no judge runs systematically against the panel on the treatment effect. Per-judge mean differences are also directionally aligned: four of five judges score Flat > Hierarchical on Writing Clarity (range +0.026 to +0.178, with GLM-5 and Qwen-3.5 significant at $p < 0.05$), and all five do so on Utility (range +0.047 to +0.120, with

Table 5. Pearson Correlations of Paired (Flat − Hierarchical) Differences Across Judges (N = 43)

| Judge | GLM-5 | GPT-5.4 | Gemini-3.1-Pro | Mistral-Large-3 | Qwen-3.5-122B |
|---|---|---|---|---|---|
| *Panel A: Writing Clarity* | | | | | |
| GLM-5 | 1.00 | 0.36 | 0.44 | 0.13 | 0.42 |
| GPT-5.4 | 0.36 | 1.00 | 0.32 | 0.10 | 0.26 |
| Gemini-3.1-Pro | 0.44 | 0.32 | 1.00 | 0.15 | 0.46 |
| Mistral-Large-3 | 0.13 | 0.10 | 0.15 | 1.00 | 0.23 |
| Qwen-3.5-122B | 0.42 | 0.26 | 0.46 | 0.23 | 1.00 |
| *Panel B: Utility* | | | | | |
| GLM-5 | 1.00 | 0.49 | 0.60 | 0.42 | 0.14 |
| GPT-5.4 | 0.49 | 1.00 | 0.31 | 0.20 | 0.11 |
| Gemini-3.1-Pro | 0.60 | 0.31 | 1.00 | 0.37 | 0.15 |
| Mistral-Large-3 | 0.42 | 0.20 | 0.37 | 1.00 | 0.09 |
| Qwen-3.5-122B | 0.14 | 0.11 | 0.15 | 0.09 | 1.00 |

Each cell is the Pearson correlation across 43 products of the paired difference (judge $j$'s Flat score minus judge $j$'s Hierarchical score) with the corresponding difference for judge $j'$. Diagonal is 1.00 by construction. Mean off-diagonal correlation: 0.29 (range 0.10–0.46) for Writing Clarity; 0.29 (range 0.09–0.60) for Utility.

GPT-5.4 and Mistral-Large-3 significant at $p < 0.05$). The directional consistency that the paired design needs is therefore present even where absolute-level Krippendorff's alpha is low: the low alpha is driven primarily by between-judge calibration offsets, which the paired design cancels by taking within-product differences.

## E Construct Validity: Inter-Dimension Correlations and Rubric Anchors

This appendix supports the interpretive caveat in the evaluation subsection of the main paper that the mapping between rubric dimensions and the exploitation/exploration constructs is an interpretive assumption of this study. We provide two pieces of evidence: a Pearson correlation matrix across the six per-dimension scores (Table 6), and the verbatim scoring rubrics issued to the LLM judges (Tables 7–8).

Table 6. Pearson Inter-Dimension Correlations (N = 86 observations, panel-mean scores)

| | Struct | Coher | Concise | Action | RCA | Strat |
|---|---|---|---|---|---|---|
| Structure | 1.00 | 0.21 | 0.07 | 0.07 | −0.16 | −0.04 |
| Coherence | 0.21 | 1.00 | 0.15 | −0.02 | 0.04 | −0.01 |
| Conciseness | 0.07 | 0.15 | 1.00 | −0.04 | 0.01 | 0.14 |
| Actionability | 0.07 | −0.02 | −0.04 | 1.00 | −0.11 | 0.09 |
| Root Cause Analysis | −0.16 | 0.04 | 0.01 | −0.11 | 1.00 | 0.06 |
| Strategic Depth | −0.04 | −0.01 | 0.14 | 0.09 | 0.06 | 1.00 |

Stacked across Flat and Hierarchical observations ($n$ = 86). The three Writing Clarity dimensions are weakly inter-correlated ($r$ = 0.07 to 0.21); the three Utility dimensions are weakly inter-correlated ($r = -0.11$ to 0.09); and cross-construct correlations are near zero or weakly negative. The composites are therefore averaging genuinely distinct signals rather than a tightly bundled latent construct, which strengthens the case for reporting per-dimension results alongside the composites and weakens any claim that a single per-dimension result "stands in for" the broader construct.

The construct interpretation in this study should be read with these correlations in mind. The mapping "Structure ↔ classical exploitation (belief convergence)" and "Strategic Depth ↔ exploration (divergent reasoning)" is theoretically

motivated (see the Theoretical Background and evaluation subsections of the main paper) but not empirically validated against an external instrument; the weak inter-dimension correlations mean that a finding on one dimension does not automatically transfer to claims about the full construct, and the per-dimension results in the Results section of the main paper should be read as descriptive evidence in their own right rather than as proxies for the composite.

Table 7. Writing Clarity Rubric Anchors (issued to all five LLM judges)

| Score | Anchor description |
|---|---|
| *Structure* | |
| 5 | ALL 6 required sections exist exactly as defined (Executive Summary, Specs, Feedback, Root Causes, Recommendations, Conclusion). Flawless markdown. |
| 4 | All 6 sections exist, but minor formatting issues (e.g., headers not bolded properly). |
| 3 | One section is missing entirely OR structure logic is moderately confused. |
| 2 | Two or more sections are missing. No clear separation of findings. |
| 1 | A single wall of text. No structure whatsoever. |
| *Coherence* | |
| 5 | Perfect logical flow. 0 contradictions. Findings perfectly align with the conclusion. |
| 4 | 1 minor break in logical flow, but 0 factual internal contradictions. |
| 3 | 1 major contradiction OR abrupt transitions. |
| 2 | Multiple contradictions. Hard to follow the narrative reasoning. |
| 1 | Completely incoherent. Random unrelated ideas strung together. |
| *Conciseness* | |
| 5 | Zero filler words. Every sentence provides new, vital data. Highly professional density. |
| 4 | Minor padding (1–2 sentences that repeat a previously stated point). |
| 3 | Noticeably verbose. Overly conversational tone unsuited for BI reporting. |
| 2 | Severe redundancy. The same exact point is made in 3 different sections. |
| 1 | Over 50% of the text is pure filler or rambling. |

Each judge receives a separate prompt per dimension instructing it to (i) produce a qualitative analysis first, (ii) then assign a 1–5 score against the strict anchors above. The judge role is framed as a "Senior Editor evaluating the Writing Clarity of a Business Intelligence report."

## F Hierarchical Manager Prompt (Verbatim)

The Hierarchical Manager's complete prompt is reproduced below. The Flat Manager prompt differs in three coupled respects that together implement the removal of coercive authority: loop-back authority is removed ("You do NOT have authority to reject work or force any worker to redo their task"), the workflow is fixed to a single forward pass, and the Manager is reframed as a peer that reviews the latest worker output (rather than the entire Blackboard) and may add only non-binding commentary. The full set of agent prompts is in the project repository.

```
You coordinate a BI report generation team of 4 workers:
- Researcher: Gathers product specs and customer reviews
- Analyst: Analyzes gathered data to find patterns and insights
- Writer: Compiles research into a professional BI report
- Critic: Reviews the report for accuracy and quality
```

Table 8. Utility Rubric Anchors (issued to all five LLM judges)

| Score | Anchor description |
|---|---|
| *Actionability* | |
| 5 | ALL recommendations perfectly follow the WHO, WHAT, WHEN, IMPACT framework. Highly implementable. |
| 4 | Most recommendations follow the framework, but 1 lacks a specific owner or timeline. |
| 3 | Recommendations exist but are generally vague (e.g., "improve battery life") with no concrete steps. |
| 2 | Only 1 vague recommendation exists. Extremely difficult to implement. |
| 1 | Zero actionable recommendations provided. Pure description. |
| *Root Cause Analysis* | |
| 5 | Traces at least 2 major flaws back to manufacturing, design, or systemic origins, supported by quote evidence. |
| 4 | Traces 1 major flaw back to an underlying systemic origin. |
| 3 | Identifies what is broken perfectly, but only guesses at why without evidence. |
| 2 | Lists symptoms (e.g., "screen is bad"), but never asks or answers why. |
| 1 | Ignores negative feedback entirely. No cause analysis. |
| *Strategic Depth* | |
| 5 | Connects product flaws explicitly to business impact (brand reputation, return rates, competitor comparison). |
| 4 | Mentions business impact broadly but lacks deep strategic competitive context. |
| 3 | Purely operational focus. Reviews product features but ignores business strategy impacts. |
| 2 | Extremely shallow. Just a summary of specs without any insights. |
| 1 | Misses the point of a BI report entirely. |

The judge role for Utility dimensions is framed as a "Business Strategist evaluating a BI report for practical value." The full prompt text for each dimension (including the qualitative-first protocol) is available in the project repository.

```
All workers share a Blackboard. You must review the ENTIRE sequence of events on the
Blackboard.
WORKFLOW: The standard sequence is: Researcher → Analyst → Writer → Critic → FINISH.
After each worker completes their task, you review the output and decide:
(1) PROCEED to the next worker in sequence (if output is sufficient), OR
(2) LOOP BACK to a previous worker (if a critical flaw is detected that must be rewritten)
LOOP-BACK RULES:
  • You have absolute authority to force a Loop-Back if the current output is corrupted.
  • You may loop back at most {max_loops} times total.
  • When looping back, provide SPECIFIC instructions on what MUST be fixed.
  • If max loops reached, you MUST proceed forward.
DECISION GUIDE:
  • After Researcher: Does the data include product specs AND customer reviews with quotes?
  • After Analyst: Are themes identified with evidence? Is root cause analysis present?
  • After Writer: Does the report follow the required 6-section structure?
```

- `After Critic: Has the Critic found major errors? Loop-Back. Has the Critic approved? Choose FINISH.`

## G Implementation Details

Table 9. System Architecture Components and Technologies

| Component | Technology | Key Reference |
|---|---|---|
| Orchestration | LangGraph + LangChain | ReAct paradigm [10] |
| LLM (Report Gen.) | 5 SOTA Models (Random Pool) | Multi-model robustness |
| LLM (Evaluation) | 5-Judge Panel | [24] |
| Vector Database | ChromaDB | [30] |
| Embeddings | all-MiniLM-L6-v2 | Sentence-BERT [29] |
| Data Source | Amazon Product Data | 52 collected, 43 retained, 540 reviews |

Three tools are wired to the experimental agents (the Researcher uses `get_product_specs` for metadata retrieval and `search_reviews` for semantic search with sentiment filtering; the Critic uses `verify_claim` for hallucination detection via similarity scoring). A fourth tool, `get_product_list`, exists in the codebase for interactive Streamlit use but is not bound to any agent in the experimental pipeline, since the target ASIN is supplied programmatically. The verification tool converts L2 distance to a similarity score:

$$\text{similarity}(q, d) = \frac{1}{1 + \|q - d\|_2} \quad \in [0, 1] \tag{1}$$

with thresholds: $\geq 0.7$ = VERIFIED, 0.5–0.7 = PARTIAL, $< 0.5$ = UNVERIFIED, formalised as

$$\text{status}(c) = \begin{cases} \text{UNVERIFIED} & \text{if similarity}(c) < 0.5 \\ \text{PARTIAL} & \text{if } 0.5 \leq \text{similarity}(c) < 0.7 \\ \text{VERIFIED} & \text{if similarity}(c) \geq 0.7 \end{cases} \tag{2}$$

The 0.5/0.7 cut-points were calibrated heuristically on a pilot inspection of representative claim–evidence pairs drawn from outside the analytic sample; the labels are reported to the Critic and Manager but are not used as numeric inputs to any downstream metric, limiting the impact of any specific threshold choice.

The overall score is the arithmetic mean of the three pillars:

$$\text{Final Score} = \frac{\text{Writing Clarity} + \text{Utility} + \text{Accuracy}}{3} \tag{3}$$

## H Additional Results Tables

**Interpreting the Structure null.** A non-significant difference test is not by itself evidence that the architectures are equivalent on Structure; it is consistent with at least three states (genuine equivalence, low statistical power for small effects, or a ceiling effect that compresses scores). To probe equivalence directly we ran a two one-sided t-test (TOST) on the paired Structure differences. At the conventional small-effect bound ($|d_z| = 0.20$, raw $|\Delta| < 0.046$), the TOST fails

Table 10. Writing Clarity Per-Dimension Comparison

| Dimension | Flat (M) | Hier (M) | Δ | Cohen's $d$ | paired-$t$ $p$ |
|---|---|---|---|---|---|
| Structure | 4.620 | 4.641 | −0.021 | −0.091 | 0.557 |
| Coherence | 4.563 | 4.465 | +0.098 | +0.192 | 0.220 |
| Conciseness | 4.179 | 3.981 | +0.197 | +0.441 | 0.007** |

** $p < 0.01$. Per-dimension tests reported uncorrected; Holm-Bonferroni-corrected $p$-values across the six per-dimension follow-ups (WC × 3 + UT × 3) are in Section A. *Note:* the composite Writing Clarity gap is driven entirely by Conciseness; Structure and Coherence do not differ significantly between architectures.

Table 11. Utility Per-Dimension Comparison

| Dimension | Flat (M) | Hier (M) | Δ | Cohen's $d$ | paired-$t$ $p$ |
|---|---|---|---|---|---|
| Actionability | 4.997 | 4.984 | +0.013 | +0.191 | 0.224 |
| Root Cause Analysis | 4.449 | 4.313 | +0.136 | +0.230 | 0.144 |
| Strategic Depth | 4.698 | 4.560 | +0.138 | +0.477 | 0.003** |

** $p < 0.01$. Per-dimension tests reported uncorrected; Holm-Bonferroni-corrected $p$-values are in Section A.

Table 12. Efficiency and System Overhead Comparison

| Metric | Flat (M ± SD) | Hierarchical (M ± SD) | Overhead |
|---|---|---|---|
| Total Cost (USD) | \$0.504 ± 0.041 | \$0.606 ± 0.189 | +20.2% |
| Generation Cost (USD) | \$0.252 ± 0.041 | \$0.354 ± 0.189 | +40.5% |
| Latency (seconds) | 209.6 ± 98.4 | 281.6 ± 159.3 | +34.3% |
| Total Tokens | 49,370 ± 11,254 | 74,781 ± 60,824 | +51.5% |

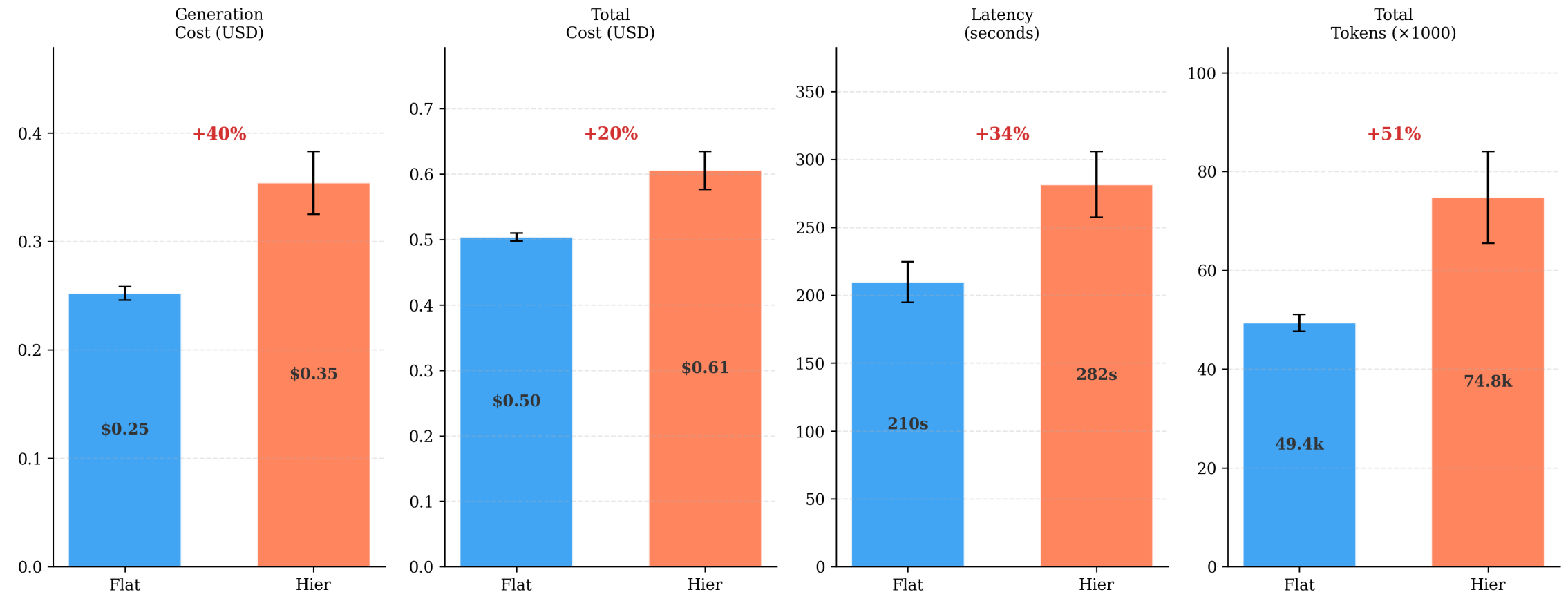


Fig. 1. Efficiency comparison between Flat and Hierarchical architectures across the four cost dimensions. Hierarchical consumes 40% more in generation cost, 20% more in total cost (generation plus evaluation), runs 34% slower, and uses 51% more tokens per report, with no corresponding quality gain on any of the three quality metrics (see Tables 10, 11, and the Results section of the main paper on specification-extraction accuracy).

Table 13. Loop Count Distribution and Quality (Hierarchical Only, N = 43)

| Loop Count | n | % | Mean WC | Mean Utility | Mean Final |
|---|---|---|---|---|---|
| 0 loops | 11 | 26% | 4.475 | 4.758 | 4.512 |
| 1 loop | 12 | 28% | 4.391 | 4.571 | 4.442 |
| 2 loops | 17 | 40% | 4.364 | 4.561 | 4.576 |
| 3 loops (blocked) | 3 | 7% | 3.790 | 4.653 | 4.373 |

*Note:* A loop count of 3 indicates the Manager requested a third revision blocked by the MAX_LOOPS = 2 guard.

Table 14. Dose–Response Robustness: Within-Hierarchical OLS for Writing Clarity

| Specification | N | $\beta$ (loop) | $p$ | $R^2$ |
|---|---|---|---|---|
| Original (loops 0–3) | 43 | −0.142 | <0.001 | 0.41 |
| Recoded (loop = 3 capped to 2) | 43 | −0.132 | 0.006 | 0.32 |
| Dropped (loop $\in \{0, 1, 2\}$ only) | 40 | −0.074 | 0.038 | 0.21 |

Each specification controls for writer model, number of reviews, and mean product rating. The loop = 3 cell contains only $n = 3$ observations (the third Manager-requested revision was blocked by the MAX_LOOPS guard); these are leverage-prone censored points. The DROP specification removes them; the RECODE specification reclassifies them as loop = 2; the ORIGINAL specification treats them at face value. The direction and significance survive in all three; the magnitude approximately halves under DROP, indicating that the censored points carry a sizeable share of the original slope.

Table 15. Paired Comparison, Subset Where Hierarchical loop_count = 0 (N = 11)

| Metric | Flat | Hier-0 | Δ (F−H) | Cohen's $d$ | paired-$t$ $p$ |
|---|---|---|---|---|---|
| Writing Clarity | 4.426 | 4.476 | −0.049 | −0.302 | 0.340 |
| Utility | 4.669 | 4.758 | −0.089 | −0.421 | 0.193 |
| Accuracy | 4.244 | 4.304 | −0.060 | −0.113 | 0.717 |

*Note:* Sign convention is standardised across all tables in this paper as Δ = Flat − Hier and Cohen's $d$ in the same direction (positive $d$ indicates Flat scores higher; negative indicates Hierarchical scores higher). In this subset Hierarchical is numerically higher than Flat on all three metrics, so all $d$ values are negative.

Table 16. Inter-Rater Reliability (Krippendorff's Alpha)

| Metric | Flat $\alpha$ | Hier $\alpha$ | Interpretation |
|---|---|---|---|
| Writing Clarity | 0.084 | 0.167 | Poor |
| Utility | 0.218 | 0.299 | Fair |

to reject ($p = 0.238$) and we cannot claim formal equivalence. At a more lenient bound ($|d_z| = 0.35$, raw $|\Delta| < 0.080$), the TOST does reject ($p = 0.048$), so we can claim equivalence only up to a small-to-medium-effect magnitude. A ceiling effect is also operating: 83.7% of Flat reports and 88.4% of Hierarchical reports score ≥ 4.5 on Structure, leaving little room for either architecture to differentiate. Together these observations support the weaker reading that Structure exhibits no detectable architectural difference at our sample size; they do not support the stronger reading that the architectures are positively shown to be equivalent on the construct most directly associated with classical exploitation.

With $N = 43$ paired observations and a two-sided paired $t$-test at $\alpha = 0.05$, our design achieves approximately 80% power to detect a medium effect size ($d = 0.44$) and approximately 61% power for a small-to-medium effect ($d = 0.35$).

The paired design further increases effective power by eliminating between-product variance, as each product serves as its own control. We do not report observed-power calculations for the realised effect sizes, since post-hoc power computed at observed effects is statistically circular [33] and adds no information beyond the $p$-value. The per-dimension follow-up tests within Writing Clarity and Utility (six dimensions) are individually underpowered for the small effects they probe and are reported as exploratory; the primary hypotheses operate at the composite level, where the a priori power calculation above applies.

## I Report Length Control

Flat reports average 1036 ± 293 words and Hierarchical reports average 1046 ± 278 words; the paired difference is −9 words ($d = -0.047$, $p = 0.760$), so the two arms are not distinguishable on length by an unconditional test. To check whether the architectural effect on each metric is nonetheless an artefact of this (statistically null) length difference, we regress the per-product paired score difference on the paired word-count difference, $\Delta\text{Score}_i = \alpha + \beta \cdot \Delta\text{Words}_i$, where $\Delta$ denotes Flat minus Hierarchical for product $i$. In this specification $\alpha$ is the architecture effect once the length difference is partialled out (the effect at equal report length), and $\beta$ is the score points per 100 words attributable to length alone.

Table 17. Architecture Effect With and Without the Length Control (Flat minus Hierarchical, N = 43 paired products)

| **Metric** | **Raw mean diff** | **Perm $p$ (raw)** | **$\alpha$ (eq. length)** | $p(\alpha)$ | **$\beta$ / 100 words** | **Share expl. by length** |
|---|---|---|---|---|---|---|
| Writing Clarity | +0.094 | 0.026 | +0.090 | 0.031 | −0.046 | 5% |
| Utility | +0.094 | 0.009 | +0.096 | 0.008 | +0.020 | −2% |
| Accuracy | −0.023 | 0.812 | −0.026 | 0.735 | −0.033 | −13% |
| Final Score | +0.055 | 0.098 | +0.053 | 0.107 | −0.019 | 3% |
| Actionability | +0.013 | 0.407 | +0.013 | 0.238 | −0.003 | 2% |
| Coherence | +0.098 | 0.227 | +0.092 | 0.240 | −0.071 | 7% |
| Conciseness | +0.197 | 0.005 | +0.193 | 0.007 | −0.044 | 2% |
| Root Cause Analysis | +0.136 | 0.146 | +0.137 | 0.145 | +0.015 | −1% |
| Strategic Depth | +0.138 | 0.003 | +0.142 | 0.002 | +0.047 | −3% |
| Structure | −0.021 | 0.560 | −0.023 | 0.509 | −0.025 | −11% |

$\alpha$ is the intercept of $\Delta\text{Score}_i = \alpha + \beta \cdot \Delta\text{Words}_i$ across the 43 paired differences (Flat minus Hierarchical): the architecture effect at equal report length. Raw mean diff and perm $p$ reproduce the unconditional paired comparison for reference. Share explained by length is the proportional shrinkage from the raw difference to $\alpha$; a negative share means $\alpha$ is larger in magnitude than the raw difference.

Table 18. Length vs Score Correlations (Pooled, N = 86)

| **Metric** | **Pearson $r$** | $p$ |
|---|---|---|
| Writing Clarity | −0.241 | 0.026 |
| Utility | +0.080 | 0.465 |
| Accuracy | +0.024 | 0.826 |
| Final Score | −0.027 | 0.802 |
| Actionability | +0.074 | 0.501 |
| Coherence | −0.001 | 0.996 |
| Conciseness | −0.459 | <0.001 |
| Root Cause Analysis | −0.015 | 0.892 |
| Strategic Depth | +0.195 | 0.073 |
| Structure | +0.086 | 0.433 |

Report length correlates negatively with Writing Clarity ($r = -0.241$, $p = 0.026$) and, most strongly, with Conciseness ($r = -0.459$, $p < 0.001$); the remaining eight metrics show no significant correlation with length.

Within the hierarchical arm, the first-stage relationship between the Manager's loop count and final report length is positive: each additional loop is associated with +105 words ($p = 0.021$). Table 19 shows the loop-count coefficient on Writing Clarity and Utility with and without the length covariate.

Table 19. Hierarchical Arm: Loop-Count Coefficient With and Without the Length Covariate (N = 43)

| Outcome | $\beta$(loops) alone | $p$ | $\beta$(loops) with length | $p$ | $\beta$(words/100) | $p$ |
|---|---|---|---|---|---|---|
| Writing Clarity | −0.126 | <0.001 | −0.102 | 0.008 | −0.023 | 0.065 |
| Utility | −0.064 | 0.067 | −0.086 | 0.019 | +0.021 | 0.078 |

The $\alpha$ inference above uses classical OLS standard errors on 43 paired differences; borderline $p$-values should be read with the same caution applied elsewhere in this supplement. The conciseness rubric partly builds report length into its scoring by design (see the Conciseness rubric anchors in Section E), so the Conciseness row in Table 17 should be read alongside the length-independent dimensions rather than in isolation.

## J Judge Sensitivity: Per-Judge and Leave-One-Judge-Out

This appendix decomposes the panel-level paired tests reported in the main paper and in Section B into per-judge effects and leave-one-judge-out (LOJO) robustness checks.

Table 20. Per-Judge Paired Effects: Writing Clarity (N = 43)

| Judge | Mean diff (F–H) | Cohen's $d$ | $t$-test $p$ | Perm $p$ |
|---|---|---|---|---|
| GPT-5.4 | −0.039 | −0.136 | 0.377 | 0.437 |
| Gemini-3.1-Pro | +0.148 | +0.294 | 0.060 | 0.064 |
| Qwen-3.5-122B | +0.179 | +0.363 | 0.022 | 0.023 |
| GLM-5 | +0.148 | +0.375 | 0.018 | 0.015 |
| Mistral-Large-3 | +0.026 | +0.068 | 0.659 | 0.660 |

Table 21. Per-Judge Paired Effects: Utility (N = 43)

| Judge | Mean diff (F–H) | Cohen's $d$ | $t$-test $p$ | Perm $p$ |
|---|---|---|---|---|
| GPT-5.4 | +0.091 | +0.307 | 0.050 | 0.065 |
| Gemini-3.1-Pro | +0.120 | +0.262 | 0.093 | 0.098 |
| Qwen-3.5-122B | +0.046 | +0.249 | 0.109 | 0.173 |
| GLM-5 | +0.116 | +0.293 | 0.061 | 0.068 |
| Mistral-Large-3 | +0.090 | +0.320 | 0.042 | 0.042 |

Per-judge tests at N = 43 are individually underpowered; the recommended reading is signs and effect sizes rather than $p$-values alone. Table 20 and Table 21 show that most judges individually point in the same direction as the panel-level result. Tables 22 and 23 are the reviewer-facing robustness check: the headline result stays significant, or close to it, whichever judge is dropped. Tables 24 and 25 repeat both views for the two per-dimension results that carry the composite-level story (Strategic Depth for Utility, Conciseness for Writing Clarity).

Table 22. Leave-One-Judge-Out: Writing Clarity (4-Judge Panel Mean, N = 43)

| Judge dropped | Mean diff | Cohen's $d$ | $t$-test $p$ | Perm $p$ |
|---|---|---|---|---|
| GPT-5.4 | +0.125 | +0.403 | 0.012 | 0.007 |
| Gemini-3.1-Pro | +0.078 | +0.301 | 0.055 | 0.054 |
| Qwen-3.5-122B | +0.071 | +0.271 | 0.083 | 0.083 |
| GLM-5 | +0.078 | +0.277 | 0.076 | 0.075 |
| Mistral-Large-3 | +0.109 | +0.351 | 0.026 | 0.020 |

Table 23. Leave-One-Judge-Out: Utility (4-Judge Panel Mean, N = 43)

| Judge dropped | Mean diff | Cohen's $d$ | $t$-test $p$ | Perm $p$ |
|---|---|---|---|---|
| GPT-5.4 | +0.093 | +0.384 | 0.016 | 0.016 |
| Gemini-3.1-Pro | +0.086 | +0.429 | 0.007 | 0.008 |
| Qwen-3.5-122B | +0.104 | +0.386 | 0.015 | 0.015 |
| GLM-5 | +0.087 | +0.425 | 0.008 | 0.008 |
| Mistral-Large-3 | +0.093 | +0.381 | 0.016 | 0.017 |

Table 24. Strategic Depth: Per-Judge and Leave-One-Judge-Out

| Analysis | Mean diff | Cohen's $d$ | $t$-test $p$ | Perm $p$ |
|---|---|---|---|---|
| Judge GPT-5.4 (N = 43) | +0.158 | +0.278 | 0.075 | 0.119 |
| Judge Gemini-3.1-Pro (N = 43) | +0.128 | +0.352 | 0.026 | 0.032 |
| Judge Qwen-3.5-122B (N = 41) | +0.098 | +0.224 | 0.160 | 0.289 |
| Judge GLM-5 (N = 42) | +0.238 | +0.406 | 0.012 | 0.018 |
| Judge Mistral-Large-3 (N = 43) | +0.100 | +0.257 | 0.100 | 0.111 |
| LOJO drop GPT-5.4 (N = 43) | +0.133 | +0.459 | 0.004 | 0.004 |
| LOJO drop Gemini-3.1-Pro (N = 43) | +0.138 | +0.413 | 0.010 | 0.010 |
| LOJO drop Qwen-3.5-122B (N = 43) | +0.153 | +0.498 | 0.002 | 0.002 |
| LOJO drop GLM-5 (N = 43) | +0.116 | +0.413 | 0.010 | 0.009 |
| LOJO drop Mistral-Large-3 (N = 43) | +0.149 | +0.458 | 0.004 | 0.005 |

Table 25. Conciseness: Per-Judge and Leave-One-Judge-Out

| Analysis | Mean diff | Cohen's $d$ | $t$-test $p$ | Perm $p$ |
|---|---|---|---|---|
| Judge GPT-5.4 (N = 43) | +0.070 | +0.271 | 0.083 | 0.250 |
| Judge Gemini-3.1-Pro (N = 43) | +0.256 | +0.244 | 0.117 | 0.145 |
| Judge Qwen-3.5-122B (N = 42) | +0.381 | +0.432 | 0.008 | 0.011 |
| Judge GLM-5 (N = 43) | +0.279 | +0.380 | 0.017 | 0.022 |
| Judge Mistral-Large-3 (N = 43) | +0.016 | +0.024 | 0.875 | 0.893 |
| LOJO drop GPT-5.4 (N = 43) | +0.229 | +0.429 | 0.007 | 0.007 |
| LOJO drop Gemini-3.1-Pro (N = 43) | +0.182 | +0.434 | 0.007 | 0.006 |
| LOJO drop Qwen-3.5-122B (N = 43) | +0.155 | +0.367 | 0.021 | 0.021 |
| LOJO drop GLM-5 (N = 43) | +0.177 | +0.376 | 0.018 | 0.018 |
| LOJO drop Mistral-Large-3 (N = 43) | +0.244 | +0.457 | 0.005 | 0.004 |

Dimension-level per-judge entries are occasionally missing in the evaluation files (12 of 2,580 judge × dimension × report cells), which is why some per-judge rows above have N = 41 or 42 rather than 43.

## K Repaired Specification-Accuracy Metric

The repaired metric scores recall and precision over the non-null ground-truth specifications for each product (5 to 7 of: ram, storage, gpu, cpu, display size, refresh rate, price, drawn from a title-preferred extraction of the product metadata). Recall is correct assertions divided by non-null ground-truth specs; precision is correct assertions divided by asserted specs. Exact hardware variants are required (e.g. RTX 4060 is not RTX 4050); a family-only assertion (e.g. a bare "Ryzen 7") counts as asserted, is flagged partial, and is scored correct only when nothing in the report contradicts the ground truth. Numeric specifications are matched numerically: price within 1 percent, and storage by marketing size class (1TB = 1000GB = 1024GB, 512GB = 500GB). This replaces the original metric used elsewhere in this paper, which scored only precision over asserted specs, imputed a neutral 3.0 when nothing was asserted, used loose substring matching, and never penalised omissions.

Table 26. Per-Arm Means, Repaired Accuracy Metric (N = 43 per Arm)

| Arm | Recall mean (sd) | Precision mean (sd) | Asserted | Correct | Partial | Omitted | Incorrect |
|---|---|---|---|---|---|---|---|
| Flat | 1.000 (0.000) | 1.000 (0.000) | 6.88 | 6.88 | 0.02 | 0.00 | 0.00 |
| Hierarchical | 0.997 (0.022) | 0.997 (0.022) | 6.88 | 6.86 | 0.02 | 0.00 | 0.02 |

Counts are per-report means over the non-null ground-truth specs. 0 of 86 reports asserted zero scoreable specs.

Table 27. Paired Flat-Minus-Hierarchical Tests, Repaired Accuracy Metric (N = 43 Product Pairs)

| Metric | Mean diff | $t$ | df | $p$ (paired $t$) | Cohen $d_z$ | $p$ (sign-flip perm.) |
|---|---|---|---|---|---|---|
| Recall | +0.003 | 1.00 | 42 | 0.3230 | 0.15 | 1.0000 |
| Precision | +0.003 | 1.00 | 42 | 0.3230 | 0.15 | 1.0000 |

Sign-flip permutation test: 100,000 draws, seed 43.

The new recall and precision metrics correlate weakly with the original accuracy score used elsewhere in this paper ($r = 0.151$, $p = 0.1664$, $n = 86$, identical for recall and precision since the two coincide almost everywhere in this sample). This correlation is not informative on its own: the new metrics are near-constant (85 of 86 reports score recall 1.0), so it is driven by a handful of points. The substantive finding is the ceiling itself: the original score varied from 3.0 to 5.0 on reports whose checkable specification content is, on the repaired metric, identical, i.e. the original variance was judge noise rather than a real difference in specification accuracy.

Table 28. Ten Largest Disagreements Between the New (Recall) and Original Accuracy Scores

| Report | Orig | Recall | Prec | Reason |
|---|---|---|---|---|
| B0CG11RG5C hierarchical | 3.0 | 1.00 | 1.00 | all 7 specs asserted and correct; original judge: Matched 2/4 specs. Mismatches: gpu: 'RTX 3060' vs 'RTX 4060'; ram: '8GB' vs '32GB RAM' |

Disagreement is $|(\text{original score} - 1)/4 - \text{recall}|$, both on a 0–1 scale (the original metric is 1–5; recall is the repaired headline metric).

*Table 28 continued*

| Report | Orig | Recall | Prec | Reason |
|---|---|---|---|---|
| B0FB19DNJW flat | 3.0 | 1.00 | 1.00 | all 7 specs asserted and correct; original judge: Matched 1/2 specs. Mismatches: gpu: ‘RTX 3050’ (not in ground truth) |
| B0FB19DNJW hierarchical | 3.0 | 1.00 | 1.00 | all 7 specs asserted and correct; original judge: Matched 1/2 specs. Mismatches: gpu: ‘RTX3060’ (not in ground truth) |
| B0BV7XQ9V9 flat | 3.7 | 1.00 | 1.00 | all 7 specs asserted and correct; original judge: Matched 2/3 specs. Mismatches: ram: ‘8GB’ vs ‘16GB DDR5’ |
| B0BXFKXG6N flat | 3.7 | 1.00 | 1.00 | all 7 specs asserted and correct; original judge: Matched 2/3 specs. Mismatches: ram: ‘8GB’ vs ‘16GB DDR5 RAM’ |
| B0BXFKXG6N hierarchical | 3.7 | 1.00 | 1.00 | all 7 specs asserted and correct; original judge: Matched 2/3 specs. Mismatches: ram: ‘8GB’ vs ‘16GB DDR5 RAM’ |
| B0C4YX39T7 flat | 3.7 | 1.00 | 1.00 | all 7 specs asserted and correct; original judge: Matched 2/3 specs. Mismatches: ram: ‘6GB’ vs ‘8GB’ |
| B0CCXHXZVP flat | 3.7 | 1.00 | 1.00 | all 7 specs asserted and correct; original judge: Matched 2/3 specs. Mismatches: ram: ‘16GB RAM’ vs ‘8GB RAM’ |
| B0CFRVSTGL flat | 3.7 | 1.00 | 1.00 | all 7 specs asserted and correct; original judge: Matched 2/3 specs. Mismatches: ram: ‘4GB’ vs ‘16GB DDR5 RAM’ |
| B0CFRVSTGL hierarchical | 3.7 | 1.00 | 1.00 | all 7 specs asserted and correct; original judge: Matched 2/3 specs. Mismatches: ram: ‘4GB’ vs ‘16GB DDR5 RAM’ |

The ground-truth parsing and report-assertion extraction rules added during validation of the repaired metric:

(1) A GB token immediately after a GPU model is read as VRAM only when it is small (≤ 16GB); otherwise storage tokens such as “1TB SSD” would be lost.
(2) Left-side RAM/Storage labels require a colon (e.g. “Memory:32GB”); without it, a token such as “512GB” inside “16GB RAM 512GB SSD” could be wrongly classified as RAM.
(3) The GPU regex tolerates trademark glyphs (e.g. “GeForce RTX(TM) 4060”).
(4) Family-only CPU ground truth (e.g. “13th Gen Intel Core i7”) is kept with a family-only flag; report assertions are scored against family and generation only.
(5) Exclusions: chassis maxima (“Up to 64GB”), bundled items (USB flash drive, docking station, portable SSD), and comparison contexts (“(>RTX 4060)”, “Beats i7-12650H”).

(6) One manual override, for a report whose description text describes a different laptop model: the description is ignored, the display size is taken from the title, and refresh rate is left null.
(7) Vendor word order (e.g. "Intel Core i5 13th Gen 13420H") counts as the full model i5-13420H.
(8) A four-digit GPU family assertion (e.g. "RTX 4050 Series Laptop GPU") counts as a full assertion; a two-digit form ("RTX 40 Series") counts only as a partial family assertion.
(9) A spec label anchors a line only in label style (followed by a colon or table pipe, allowing bold markers, padding, and parenthetical aliases such as "Memory (RAM):"); prose mentions (e.g. a price-point aside or an upsell recommendation in an executive summary) no longer hijack the assertion.
(10) Review quotes are stripped before scanning (straight and curly quotes; inch marks after digits are not treated as quote delimiters), so customer-reported discrepancies do not count as report assertions; discrepancy discussions outside the spec table were individually checked, and in every case the report itself separately asserts the listing spec.

## L Draft-Level Text Measures: Hedging, Lexical Diversity, and Revision Overlap

**Hedging density.** Hedges per 1000 whitespace-delimited words, using a fixed lexicon of 51 single-word items and 18 multiword phrases, matched case-insensitively on word boundaries; multiword phrases are matched first and masked so their component words are not double-counted. The core lexicon is: may, might, could, possibly, possible, likely, unlikely, perhaps, appears, appear, seems, seem, suggests, suggest, potentially, presumably, arguably, somewhat, relatively, generally, typically, tends, tend, roughly, approximately, "it is possible", "may be". Documented additions cover three families: inflectional variants the core list omits (appeared, seemed, suggested, suggesting, tended, probable, probably); further epistemic markers of the same class (apparently, seemingly, reportedly, conceivably, presumable, maybe, indicate, indicates, indicated); and approximators and downtoners matching the core "somewhat / relatively / roughly" (fairly, slightly, moderately, largely, mostly, often, sometimes, occasionally, marginally, partially).

**Lexical diversity.** MTLD, computed as specified by McCarthy and Jarvis (2010): a forward pass over lowercased alphabetic tokens accumulates a factor each time the running type-token ratio falls to or below 0.72, with a fractional partial factor for the trailing run computed as $(1-\mathrm{TTR}_{\mathrm{final}})/(1-0.72)$ and MTLD = tokens/factors; the same computation is run in reverse and the two values averaged. A companion measure, moving-average type-token ratio with window 100 (MATTR(100)), is reported alongside because it is insensitive to text length by construction.

**Draft-to-draft overlap.** Jaccard similarity over the set of content-word types (lowercased alphabetic tokens with a 135-item closed-class stopword list removed), and an LCS-based similarity: the length of the longest common subsequence of the two drafts' word sequences divided by the length of the later draft.

Table 29. Draft-Extraction Validation

| **Validation check** | **Result** |
|---|---|
| Flat runs whose final extracted draft matches `report_text` | 43/43 (43 exact) |
| Hierarchical runs whose final extracted draft matches `report_text` | 43/43 (42 exact, 1 prefix) |
| Extracted drafts that are markdown-report shaped (≥ 800 chars, ≥ 3 headings) | 120/120 |
| Flat runs with exactly one draft | 43/43 |

The one non-exact case is the hierarchical run for ASIN B0FLPY45RX (loop_count 3): its final Writer draft (923 words) is a strict prefix of `report_text` (1499 words), because the run ended on an Analyst turn and the harness appended a trailing "ANALYST INSIGHTS" block from that Analyst output. The Writer draft itself is recovered correctly; the extra 576 words are Analyst text, not part of the draft.

Table 30. Drafts per Run vs Loop Count (Hierarchical, N = 43)

| loop_count | 1 draft | 2 drafts | 3 drafts | Total runs |
|---|---|---|---|---|
| 0 | 11 | 0 | 0 | 11 |
| 1 | 0 | 12 | 0 | 12 |
| 2 | 0 | 15 | 2 | 17 |
| 3 | 1 | 1 | 1 | 3 |
| Total | 12 | 28 | 3 | 43 |

Total hierarchical Writer drafts: 77. Runs with 2 or more drafts: 31.

Table 31. Observed Turn Sequences by Loop Count (Hierarchical, N = 43)

| loop_count | runs | Turn sequence |
|---|---|---|
| 0 | 8 | Analyst > Writer > Critic |
| 0 | 3 | Writer > Critic |
| 1 | 10 | Analyst > Writer > Critic > Writer |
| 1 | 2 | Writer > Critic > Writer |
| 2 | 14 | Analyst > Writer > Critic > Writer > Critic |
| 2 | 2 | Analyst > Writer > Critic > Writer > Writer |
| 2 | 1 | Writer > Critic > Critic > Writer |
| 3 | 1 | Analyst > Writer > Critic > Analyst |
| 3 | 1 | Analyst > Writer > Critic > Writer > Analyst |
| 3 | 1 | Analyst > Writer > Critic > Writer > Writer |

`loop_count` does not equal the number of Writer rewrites; it counts role turns beyond the initial pass, in which the Manager may re-delegate to the Writer, the Critic, or the Analyst. For 40 of the 43 hierarchical runs, loop_count equals the number of role turns minus 3 (if the run had an Analyst turn) or minus 2 (otherwise); the three loop_count = 3 runs are the exceptions, each ending on a turn that produced no new draft, consistent with a Manager delegation hitting the MAX_LOOPS cap before producing output. All 43 flat runs share the identical turn sequence Analyst > Writer > Critic.

Table 32. Final-Report Descriptives by Arm (N = 86, 43 Matched Pairs)

| Arm | Word count | Hedge tokens (raw) | Hedges/ 1000 words | MTLD | MATTR(100) |
|---|---|---|---|---|---|
| Flat (N = 43) | 1036.4 (293.4) | 3.88 (4.58) | 3.298 (2.844) | 189.07 (26.66) | 0.818 (0.014) |
| Hierarchical (N = 43) | 1045.7 (277.7) | 6.05 (6.78) | 5.032 (3.943) | 186.42 (26.83) | 0.814 (0.019) |

Means with standard deviations in parentheses.

Not one of the 34 transitions is a substantial rewrite by any reasonable threshold; the median revised draft retains 88% of its predecessor as a common subsequence and 85% of its content-word vocabulary while growing by a median of

Table 33. Paired Tests, Flat Minus Hierarchical (N = 43 Pairs)

| Measure | Flat | Hier. | Diff | $t(42)$ | $p$ ($t$) | $d_z$ |
|---|---|---|---|---|---|---|
| Hedges per 1000 words | 3.298 | 5.032 | −1.734 | −3.981 | <0.001 | −0.607 |
| MTLD | 189.07 | 186.42 | 2.652 | 0.451 | 0.654 | 0.069 |
| MATTR(100) | 0.818 | 0.814 | 0.004 | 1.225 | 0.227 | 0.187 |
| Word count | 1036.4 | 1045.7 | −9.26 | −0.308 | 0.760 | −0.047 |

Wilcoxon and permutation tests agree in sign and significance with the paired $t$-test in every row. Mechanism verdict: hedging **supports** the iterative-revision account (Hierarchical reports hedge about 53% more per 1000 words); lexical diversity is **neutral**, since a vocabulary-collapse reading of Degeneration-of-Thought would predict a detectable MTLD/MATTR difference and none is found.

Table 34. Top 15 Hedge Items by Combined Frequency (Raw Counts Summed over 43 Reports per Arm)

| Item | Flat | Hierarchical | Difference |
|---|---|---|---|
| likely | 38 | 43 | +5 |
| possible | 12 | 31 | +19 |
| may | 13 | 29 | +16 |
| indicates | 13 | 14 | +1 |
| suggests | 7 | 20 | +13 |
| appears | 11 | 11 | 0 |
| could | 5 | 11 | +6 |
| indicate | 8 | 8 | 0 |
| appear | 6 | 10 | +4 |
| suggest | 8 | 7 | −1 |
| potentially | 3 | 9 | +6 |
| may be | 7 | 3 | −4 |
| slightly | 3 | 6 | +3 |
| generally | 3 | 4 | +1 |
| apparently | 5 | 2 | −3 |
| **All items** | **167** | **260** | **+93** |

Distinct hedge items used: 31 in the flat arm, 40 in the hierarchical arm. The gap is spread across the modal and evidential core (possible, may, suggests, could, potentially) rather than produced by one idiosyncratic term.

Table 35. Arm Gap Split by Whether the Manager Forced a Revision

| Subgroup | Measure | Flat | Hier. | Diff | $d_z$ | $p$ ($t$) | $p$ (perm) |
|---|---|---|---|---|---|---|---|
| loop_count = 0 (N = 11) | Hedges/1000 words | 1.999 | 2.899 | −0.900 | −0.487 | 0.137 | 0.125 |
| loop_count = 0 (N = 11) | MTLD | 188.92 | 185.25 | 3.677 | 0.119 | 0.702 | 0.697 |
| loop_count = 0 (N = 11) | Word count | 922.4 | 950.6 | −28.27 | −0.170 | 0.585 | 0.649 |
| loop_count ≥ 1 (N = 32) | Hedges/1000 words | 3.744 | 5.765 | −2.021 | −0.652 | <0.001 | <0.001 |
| loop_count ≥ 1 (N = 32) | MTLD | 189.13 | 186.83 | 2.300 | 0.056 | 0.755 | 0.757 |
| loop_count ≥ 1 (N = 32) | Word count | 1075.6 | 1078.3 | −2.72 | −0.013 | 0.942 | 0.944 |

The arm-level hedging gap is concentrated in the pairs where the Manager intervened. The N = 11 no-revision subgroup is underpowered on its own; the properly powered version of this comparison is the hierarchical-draft-1-vs-flat-report test below, which uses all 43 pairs.

Table 36. Hierarchical Arm: OLS on loop_count, With and Without Word Count (N = 43, Classical Homoskedastic SEs)

| Outcome | Model | loop_count | SE | $p$ | word_count | SE | $p$ | $R^2$ |
|---|---|---|---|---|---|---|---|---|
| Hedges/1000 words | no covariate | 1.738 | 0.601 | 0.006 | | | | 0.169 |
| Hedges/1000 words | + word count | 0.738 | 0.470 | 0.124 | 0.0095 | 0.0016 | <0.001 | 0.566 |
| MTLD | no covariate | −3.365 | 4.455 | 0.454 | | | | 0.014 |
| MTLD | + word count | −2.015 | 4.781 | 0.676 | −0.0129 | 0.0161 | 0.427 | 0.029 |
| MATTR(100) | no covariate | −0.0039 | 0.0031 | 0.213 | | | | 0.038 |
| MATTR(100) | + word count | −0.0024 | 0.0032 | 0.456 | −0.0000137 | 0.0000109 | 0.218 | 0.074 |
| Word count | no covariate | 104.72 | 43.45 | 0.021 | | | | 0.124 |

Standard errors are classical; with N = 43 and a discrete four-valued regressor this is a small model, and the coefficients are descriptive dose slopes rather than causal estimates, since loop_count is Manager-assigned. The word-count row reproduces the established first-stage result (+104.7 words per loop, $p = 0.021$). The loop_count slope on hedging falls from 1.738 to 0.738 and loses significance once word count enters; since hedging is already expressed per 1000 words this is not a length artefact of the measure, but length and loop_count are collinear in this arm by construction, so this between-run coefficient should be treated as fragile relative to the within-run draft comparisons below.

Table 37. Draft-to-Draft Content Overlap (Hierarchical Arm)

| Transition | n | Jaccard mean (sd) | Jaccard median | LCS sim mean (sd) | LCS sim median | LCS sim min | Δwords mean (median) |
|---|---|---|---|---|---|---|---|
| Draft 1 to draft 2 | 31 | 0.827 (0.139) | 0.849 | 0.836 (0.138) | 0.882 | 0.510 | +38.2 (+30) |
| Draft 2 to draft 3 | 3 | 0.966 (0.030) | 0.951 | 0.974 (0.023) | 0.966 | 0.956 | −1.0 (0) |
| All transitions | 34 | 0.839 (0.139) | 0.872 | 0.849 (0.138) | 0.890 | 0.510 | +34.7 (+22) |

Table 38. Classification of Draft-to-Draft Transitions

| Classification | Count |
|---|---|
| Transitions with LCS similarity ≥ 0.60 (light edit or append) | 30/34 |
| Transitions with LCS similarity < 0.30 (substantial rewrite) | 0/34 |
| Transitions where the later draft is longer | 25/34 |
| Transitions where the later draft is byte-identical after whitespace normalisation | 1/34 |

30 words. Mechanism verdict: **partly supports, partly undercuts**. Revision does not look like reconsideration (the Writer edits at the margin and appends), which undercuts the strong reading of Degeneration-of-Thought as progressive drift into worse content; what changes across drafts is stance and length, not substance.

Table 39. Hedging and Lexical Diversity by Draft Index (Hierarchical Arm, Unbalanced)

| Draft index | n | Words mean (sd) | Hedges/1000 words mean (sd) | MTLD mean (sd) | MATTR(100) mean (sd) |
|---|---|---|---|---|---|
| 1 | 43 | 1004.8 (263.9) | 3.568 (3.492) | 188.94 (23.62) | 0.818 (0.017) |
| 2 | 31 | 1037.9 (265.9) | 5.448 (4.064) | 187.46 (23.71) | 0.815 (0.018) |
| 3 | 3 | 899.7 (103.0) | 5.976 (3.036) | 200.83 (63.10) | 0.811 (0.029) |
| Flat single draft | 43 | 1036.4 (293.4) | 3.298 (2.844) | 189.07 (26.66) | 0.818 (0.014) |

Iterative revision under a Manager leaves a specific, measurable trace in the text, and it is a stance trace rather than a content trace: the same Writer, on the same product, with the same underlying model, returns a draft that hedges roughly 56% more per 1000 words ($d_z = 0.833$, $N = 31$) and runs about 38 words longer, while preserving roughly 88% of its own previous text as a common subsequence. The hierarchical Writer's first draft is statistically indistinguishable

Table 40. Draft 1 vs Draft 2, Paired (N = 31 Runs)

| **Measure** | **Draft 1** | **Draft 2** | **Diff** | $t$**(30)** | $p$ ($t$) | $d_z$ |
|---|---|---|---|---|---|---|
| Hedges per 1000 words | 3.498 | 5.448 | +1.950 | 4.636 | <0.001 | 0.833 |
| MTLD | 190.33 | 187.46 | −2.868 | −1.157 | 0.256 | −0.208 |
| MATTR(100) | 0.820 | 0.815 | −0.005 | −2.828 | 0.008 | −0.508 |
| Word count | 999.7 | 1037.9 | +38.16 | 2.878 | 0.007 | 0.517 |

Mechanism verdict: **supports**, and it is the cleanest evidence in the set, since product, role assignment and the underlying model are held fixed and only the revision act varies. Lexical diversity gives a weaker, mixed signal: MTLD does not move ($d_z = -0.208$, $p = 0.256$) while MATTR falls slightly but reliably ($d_z = -0.508$, $p = 0.008$); since MTLD is the primary specified measure, the honest reading is at most a marginal effect on lexical diversity alongside a clear effect on hedging.

Table 41. Draft 2 vs Draft 3, Paired (N = 3 Runs; Reported for Completeness Only, No Inference Supported)

| **Measure** | **Draft 2** | **Draft 3** | **Diff** | $t$**(2)** | $p$ ($t$) | $d_z$ |
|---|---|---|---|---|---|---|
| Hedges per 1000 words | 5.615 | 5.976 | +0.360 | 1.145 | 0.371 | 0.661 |
| MTLD | 194.71 | 200.83 | +6.114 | 0.395 | 0.731 | 0.228 |
| MATTR(100) | 0.814 | 0.811 | −0.003 | −1.909 | 0.197 | −1.102 |
| Word count | 900.7 | 899.7 | −1.00 | −0.084 | 0.940 | −0.049 |

N = 3 supports no inference; the permutation test has only 8 distinct sign assignments. Reported for completeness. The three runs reaching a third draft, word count then hedges per 1000 words then MTLD from draft 1 to draft 3: B0C4YX39T7 (820 > 852 > 830 words; 1.22 > 3.52 > 3.61 hedges/1000; 232.8 > 232.8 > 268.2 MTLD), B0F195W823 (964 > 1018 > 1018 words; 3.11 > 4.91 > 4.91 hedges/1000; 211.0 > 191.1 > 191.1 MTLD), B0F538NQPH (823 > 832 > 851 words; 2.43 > 8.41 > 9.40 hedges/1000; 192.7 > 160.2 > 143.2 MTLD).

Table 42. Hierarchical Draft 1 vs the Flat Report, Paired (Same 43 Products)

| **Measure** | **Hier. draft 1** | **Flat report** | **Diff** | $t$**(42)** | $p$ ($t$) | $d_z$ |
|---|---|---|---|---|---|---|
| Hedges per 1000 words | 3.568 | 3.298 | +0.270 | 0.750 | 0.457 | 0.114 |
| MTLD | 188.94 | 189.07 | −0.130 | −0.026 | 0.979 | −0.004 |
| MATTR(100) | 0.818 | 0.818 | +0.001 | 0.193 | 0.848 | 0.029 |
| Word count | 1004.8 | 1036.4 | −31.58 | −1.305 | 0.199 | −0.199 |

Before revision the two arms' reports are textually indistinguishable on every measure here, hedging included ($d_z = 0.114$, $p = 0.457$). Mechanism verdict: **supports**, and it is the load-bearing result; the full −1.734 hedging gap in the final reports opens entirely between the hierarchical Writer's first draft and its last, so the hedging difference is an effect of being told to revise, not an effect of hierarchy as a structure.

from the flat arm's report on every measure here, so the entire arm-level hedging gap is manufactured by the revision loop rather than by the hierarchy itself. This is a direct fit to sycophantic compliance (instructed to respond to criticism, the Writer softens its claims and adds qualifying material rather than reconsidering the analysis), and a poorer fit to Degeneration-of-Thought in its strong form, which implies a progressive drift into worse or more degenerate content: overlap stays high, no transition is a substantial rewrite, and lexical diversity is flat both between arms and within runs. Two caveats apply throughout this appendix. The hedge lexicon cannot distinguish well-calibrated uncertainty from evasion, so the reading of hedging as harm rests on the judge quality scores, not on the lexicon itself. And the between-run loop_count coefficients, including the +105-words-per-loop figure, are Manager-assigned and partly reflect selection into revision rather than the causal effect of revising; the within-run draft comparisons in this appendix are the estimates to lead with.

Table 43. Summary of Verdicts on the Degeneration-of-Thought / Sycophancy Mechanism

| Finding | N | Effect | Verdict |
|---|---|---|---|
| Hierarchical final reports hedge more than flat | 43 pairs | $d_z = -0.607$, $p < 0.001$ | Supports |
| Arms are identical in lexical diversity (MTLD, MATTR) | 43 pairs | $d_z = 0.069$ / $0.187$, $p = 0.654$ / $0.227$ | Neutral; rules out a vocabulary-collapse reading |
| Hierarchical draft 1 is indistinguishable from the flat report | 43 pairs | hedging $d_z = 0.114$, $p = 0.457$ | Supports (localises the effect to revision, not structure) |
| Revision raises hedging within the same run | 31 pairs | $d_z = 0.833$, $p < 0.001$ | Supports, strongest evidence |
| Revision leaves MTLD unchanged, lowers MATTR slightly | 31 pairs | $d_z = -0.208$ ($p = 0.256$) / $-0.508$ ($p = 0.008$) | Weak, mixed; at most marginal |
| Successive drafts overlap heavily, no rewrites | 34 transitions | median LCS sim 0.890, 0/34 below 0.30 | Supports “no reconsideration”; undercuts “progressive degeneration” |
| Revision lengthens the same report | 31 pairs | +38 words, $d_z = 0.517$, $p = 0.007$ | Supports the padding channel; qualifies the +105 figure |
| Arm hedging gap concentrated in revised pairs | 11 vs 32 pairs | $p = 0.137$ vs $p < 0.001$ | Supports (underpowered on its own) |
| loop_count slope on hedging is fragile to a length control | 43 | 1.738 ($p = 0.006$) to 0.738 ($p = 0.124$) | Caution; prefer the within-run test |



\end{document}